\documentclass[12pt]{article}
\usepackage{geometry,float}
\usepackage{amssymb}
\usepackage{amstext,enumerate}
\usepackage{bm}
\usepackage{graphics}
\usepackage{graphicx}
\usepackage{amsthm}
\usepackage{amsmath}
\usepackage{latexsym}
\usepackage{rotate}
\usepackage{lscape}
\usepackage{color}
\usepackage{epsfig,epsf,psfrag}
\usepackage{sectsty}
\usepackage{graphics}
\usepackage{epsfig}
\usepackage{multirow}
\usepackage{graphicx}
\usepackage{amsmath}
\usepackage{url}
\usepackage{amsthm}
\usepackage{amssymb}
\usepackage{sectsty}
\usepackage{epsfig,natbib}
\usepackage{rotate}
\usepackage{lscape}
\usepackage{setspace}
\usepackage{subcaption}
\usepackage{float}
\usepackage{graphicx}
\usepackage{mathtools}
\newtheorem{Theorem}{Theorem}

\include{mathdefn2}

\sectionfont{\centering\bf\sf\normalsize}
\subsectionfont{\sf\normalsize}

\renewcommand{\baselinestretch}{1.6} 
\newcommand{\single}{\renewcommand{\baselinestretch}{1.2}\normalsize}
\newcommand{\double}{\renewcommand{\baselinestretch}{1.63}\normalsize}

  \renewenvironment{thebibliography}[1]{
    \begin{oldthebibliography}{#1}
      \setlength{\parskip}{0ex}
      \setlength{\itemsep}{0ex}
  }
  {
    \end{oldthebibliography}
  }

\newcommand{\bea}{\begin{eqnarray*}}
\newcommand{\eea}{\end{eqnarray*}}
\newcommand{\be}{\begin{eqnarray}}
\newcommand{\ee}{\end{eqnarray}}
\newcommand{\ed}{\end{document}}

\newcommand{\btab}{\begin{tabular}}
\newcommand{\etab}{\end{tabular}}

\newcommand{\la}{\label}
\newcommand{\bi}{\begin{itemize}}
\newcommand{\ei}{\end{itemize}}
\newcommand{\bfi}{\begin{figure}}
\newcommand{\efi}{\end{figure}}
\newcommand{\ben}{\begin{enumerate}}
\newcommand{\een}{\end{enumerate}}
\newcommand{\bay}{\begin{array}}
\newcommand{\eay}{\end{array}}

\newcommand{\rs}{\vspace{-.2cm}}

\def\vs{\vspace{.5cm}}

\def\bco{\iffalse}

\def\cp{\citep}

\def\rs{\vspace{-.1in}}

\newcommand{\no}{\noindent}
\newcommand{\bc}{\begin{center}}
\newcommand{\ec}{\end{center}}

\DeclareMathOperator*{\argmin}{argmin}

\newcommand{\sm}{}
\bibpunct{(}{)}{;}{a}{}{,}

\begin{document}
\thispagestyle{empty} \single \bc {\bf \sc \Large Dynamic Modeling with Conditional Quantile Trajectories for Longitudinal Snippet Data, with Application to Cognitive Decline of Alzheimer's Patients$^{\dagger *}$}\vspace{0.15in}\\
Matthew Dawson$^1$ and  Hans-Georg M\"uller$^2$ \\
$^1$ Graduate Group in Biostatistics, University of California, Davis \\
$^2$ Department of Statistics, University of California, Davis\\
Davis, CA 95616 USA \ec \centerline{February 2017}

\vspace{0.1in} \thispagestyle{empty}
\bc{\bf \sf ABSTRACT} \ec \vspace{-.1in} \no 
\setstretch{1}
Longitudinal data are often plagued with sparsity of time points where measurements are available. The functional data analysis perspective has been shown to provide an effective and flexible approach to address this problem for the  case where measurements are sparse but their times are randomly distributed over an interval. Here we focus on a different scenario where available data can be characterized as  snippets, which are very short stretches of longitudinal measurements. For each subject the stretch of available data is much shorter than the time frame of interest, a common occurrence in accelerated longitudinal studies. An added challenge is introduced if a time proxy that is basic for usual longitudinal modeling is not available. This situation arises in the case of Alzheimer's disease and comparable scenarios,  where one is interested in time dynamics of declining performance, but the time of disease onset is unknown and the chronological age does not provide a meaningful time reference for longitudinal modeling. Our main methodological contribution is to address this problem with a novel approach. 
Key quantities for our approach are conditional quantile trajectories for monotonic processes that emerge as solutions of a dynamic system, and for which we obtain uniformly consistent estimates.  These trajectories are shown to be useful to describe processes that quantify deterioration over time, such as hippocampal volumes in Alzheimer's patients.\\

\no {KEY WORDS:\quad Accelerated longitudinal study, autonomous differential equation, uniform convergence, monotonic process, Functional data analysis, nonparametric estimation, hippocampal volume}.
\thispagestyle{empty} \vfill
\noindent \vspace{-.2cm}\rule{\textwidth}{0.5pt}\\
{\small $^\dagger$ Research supported by NSF grants DMS-1228369 and
DMS-1407852.\\ $^*$ The data used in this paper are from the Alzheimer's Disease Center at University of California Davis, supported by NIH and NIA grant P30 AG10129}

\newpage
\pagenumbering{arabic} \setcounter{page}{1} \double

\bc {\bf \sf 1.\quad INTRODUCTION}\sm \ec \rs

\no When adopting the functional approach for the analysis of longitudinal data, a common assumption is that the observations originate from a smooth underlying process. This assumption is justified, for example, when the observations correspond to biological mechanisms which are known to vary smoothly, and in this case modeling longitudinal data with functional data analysis (FDA) approaches has been highly successful \citep{brum:98, stan:98,  rice:01,  rice:04, Guo:04:1, jian:11, 
coff:14, Wang2003, Wang2005}. A common methodological challenge in longitudinal studies, however, is that many such studies lack complete and densely spaced  observations over time.  Some authors have explored problems relating to incomplete functional data \citep{Yao2005, Delaigle2013,Kraus2014, liebl:16,Delaigle2016}. These methods involve estimation of the covariance function of the underlying random functions or of transition probabilities in a Markov chain, either by pooling or stitching observed fragments. In this paper, we consider a new quantile based approach, which may be used when dealing with a severe type of sparseness that substantially differs from previous approaches.  

Some data generated in longitudinal studies exhibit an extreme form of sparseness. We refer to such data as snippet data, often originating from \emph{accelerated longitudinal designs} \cp{Galbraith2014}. Snippet data can be characterized as very short longitudinal measurements relative to the domain of interest. A design of this type is attractive to practitioners across the social and life sciences since it minimizes the length of time over which one needs to gather data for each subject; they are especially useful in situations where data collection is invasive, difficult or expensive, as is for example the case when studying Alzheimer's disease.

\begin{figure}[H]
  \begin{center}
  \begin{subfigure}[H]{.49\textwidth}
    \includegraphics[width = \textwidth, height = 2.5in]{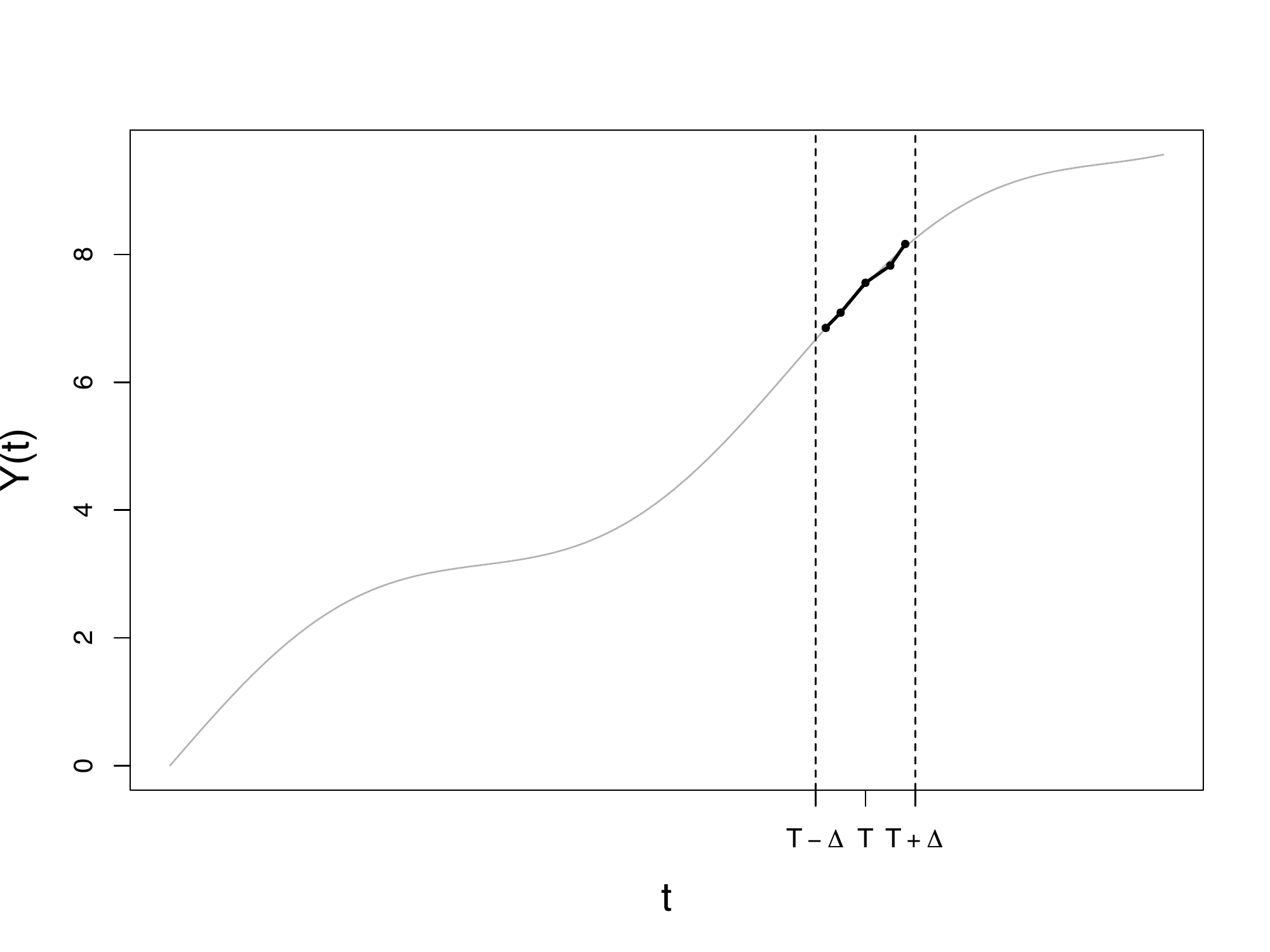}
  \end{subfigure}
  \begin{subfigure}[H]{.49\textwidth}
    \includegraphics[width = \textwidth, height = 2.5in]{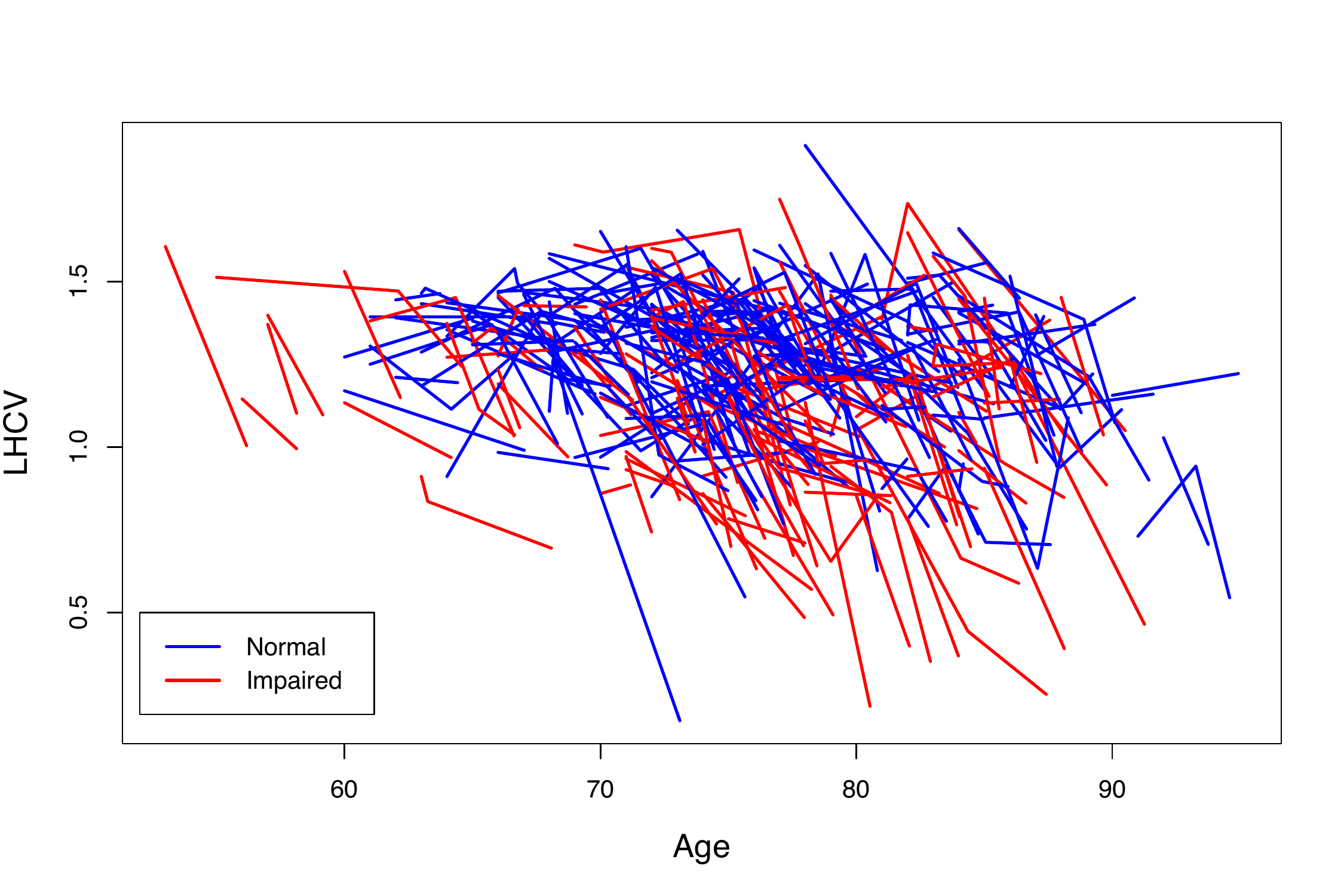}
  \end{subfigure}
    \caption{A simulated trajectory  with an example of a longitudinal snippet at time $T$ (left) and log-hippocampal volume snippet data, where age is the only available time reference (right). Subjects classified as having normal cognitive function are colored in blue, while those with impaired cognitive function are colored in red.}
    \label{fig:snippetGen}
  \end{center}
\end{figure}

Snippet data may be viewed as being  generated by observing each subject for a short window around some random time $T$. An illustration of how snippets originate is shown in the left panel of Figure \ref{fig:snippetGen}.  Of particular interest is the case where the subjects' entry times are not informative, which is often the case in studies where, for example, the time since disease onset is unknown but where this time plays a decisive role. 

Typical methods for dealing with snippet data resulting from accelerated longitidinal studies involve parametric models \citep{Raudenbush1992, acc1, acc2, acc3, acc4, acc5}. These methods do not allow the recovery of the underlying functional dynamics, which  has not been systematically studied so far. For snippet data where  an absolute time scale such as age of subjects  is not informative, functional completion methods based on stitching or pooling segments are not valid. The covariance function is also not estimable which also precludes most functional and longitudinal models, including those in \citet{Yao2005}, \citet{Kraus2014}, and \citet{liebl:16}. Figure \ref{fig:sparseVSnippet} in the Supplement clearly demonstrates why the covariance function cannot be estimated in the snippet case, even when the observed time is informative, in contrast to the usually considered sparse case where measurements are randomly located over the entire domain.

The focus of this paper is to flexibly estimate quantile dynamics of the underlying smooth process, including the difficult case where the subjects' available entry times $T_i$ such as age are not necessarily useful, a situation which arises when measuring deterioration since an unknown onset of disease or degradation in degradation experiments.  There exists substantial literature on parametric and nonparametric methods for quantile and conditional quantile estimation \citep{quant:fan, quant:hall, quant:koenker1, quant:koenker2, quant:koenker3, quant:li, quant:wei, quant:yu}. While these methods and their extensions are applicable in simpler scenarios where full curves are observed for each subject, to our knowledge estimation of quantile trajectories based on snippet data has not yet been considered. To be specific, we note that obtaining cross-sectional quantile trajectories, conditional on $T$, is not meaningful in this context since there is no guarantee of proper alignment of snippets. 

Under the scenario where $T_i$ is not necessarily useful and the underlying process is monotonic, \citet{Abramson1994} and \citet{Vittinghoff1994}  found that one can still obtain trajectory information  over time as long as information about local level and slope is known for each subject.  Specifically, Abramson and M\"uller suggested that data in this form can be viewed as bivariate observations of level and slope at some random, unobserved time. Formally, we may write  $X_i = f(T_i)$ and $Z_i = f'(T_i) + \epsilon_i$,  for $i = 1, \dots, n$, with i.i.d. noise $\epsilon_i$ satisfying $E(\epsilon) = 0$ and $E(\epsilon^2) = \sigma^2 < \infty$, where $f$ is a fixed, strictly monotonic function and one observes $(X_i,Z_i)$. The key observation that was simultaneously  made  in  \citet{Vittinghoff1994}  is that for a monotonic function, there exists a function $g$ that relates the slope to the level, i.e.,  $f'(t) = g(f(t))$,  where $g(x) = E(Z | X = x)$ and we can use the available snippet data to estimate this function using scatterplot smoothers \citep{Fan1996}.   These approaches reflect that due to the short time span of  snippet data, the available data do not carry information beyond local level and slope. To extract this local information, one can apply a simple linear least squares fit to the data in each snippet and extract slope and mean estimates. 

A key assumption in \citet{Abramson1994, Vittinghoff1994} was that one  has  measurements from a fixed function $f$ which is the target and corresponds to a population mean function.  However, in addition to the mean function,  individual dynamics are of paramount interest, for example in accelerated longitudinal studies. To  target individual dynamics,  we assume that observations to come from realizations of a stochastic process, and aim to estimate functionals of the conditional distribution of slopes, rather than only the mean function. The proposed methods combine known results for conditional quantile estimation with the underlying smoothness assumptions and dynamics that are the foundation of functional data analysis.  The result is a straightforward method for estimating conditional quantile trajectories from snippet data that is supported by theory. 

The organization of this paper is as follows. In Section 2 we introduce the proposed dynamic model, while Section 3 covers estimation procedures. Our main theoretical results are discussed in Section 4. Simulations and an application to Alzheimer's disease are discussed in Sections 5 and 6, respectively.

\bc {\bf \sf 2.\quad DYNAMIC MODELING OF THE DISTRIBUTION OF DECLINE RATES}\sm \ec \rs
\noindent { \sf 2.1. \quad Basic Model}

\no  We assume that the observed snippets are generated by an underlying random process $Y$, which is  defined on some domain $\mathcal{T}_0$  
and is  $(k + 1)$-times continuously differentiable for a $k \geq 1$. Let $\mathcal{J}$ be the range of $Y$ restricted to $\mathcal{T}_0$.  Measurements are generated by observing $Y_i(T_i)$ and $Y_i'(T_i)$ at some random and potentially unobserved subject-specific time $T_i$, where $T$ is independent of $Y$. Denote these observations as $X_i \coloneqq Y_i(T_i)$ and $Z_i \coloneqq Y_i'(T_i)$ for $i = 1, \dots, n$. If $X_i$ and $Z_i$ are not directly observable
one may use surrogates for level and slope of the snippets, which can be obtained by least squares line fits to the observed snippet data for each subject. Further assume that $T_i \sim f_T$ for some density $f_T$ on $\mathcal{T}_0$, and that $(X_i, Z_i, T_i)$ have a joint distribution and are independent of $(X_j, Z_j, T_j)$ for $i \neq j$. A core feature of our model is that the conditional distribution of the slope given the level can be expressed in a way that does not explicitly depend on $T$, which means that we may view the distribution of the rate of decline  as a function of the current level without knowledge of the observation time, i.e.,
\begin{equation}
F(z|x) = P(Y'(T) \leq z \bigm| Y(T)=x) = P(Z \leq z \bigm| X = x), 
\label{eq : model}
\end{equation}
is determined by the relationship between level and slope, while bypassing $T$.  This view is similar to assumptions made in \citet{Abramson1994} and \citet{Vittinghoff1994}, though the focus there  was on estimation of the mean, rather than the conditional distribution. It should be emphasized that our assumption does not imply that the derivative $Y'(T)$ does not depend on $T$; rather we change the frame of reference from conditioning on time to conditioning on level. Monotone processes lend themselves nicely to this perspective, noting that if $Y$ is monotonic and differentiable we have
\begin{equation}
P(Y'(T) \leq z \bigm| Y(T)) = P(Y'(Y^{-1}(Y(T))) \leq z \bigm| Y(T)),
\label{eq : monotonic}
\end{equation}
so that $T$ only comes into the conditional distribution via $Y(T)$.

For our application to Alzheimer's hippocampal volume data we find this type of model to be appropriate, and there is evidence that the distribution of decline rates conditioned on level is independent of age. For example, when segmenting the dataset  according to level and age, and comparing within each level segment the distribution of slopes $Z$ for different ages (see Figure \ref{fig:empiricalComparison} in the Supplement) we find that there is no obvious relationship. We also find that when fitting a linear model with slope as response and  level as predictor,  there is no evidence from an F test that  adding age as an additional predictor improves the regression relation,  indicating that the level is sufficient in modeling the mean slope. We obtain similar results when fitting parametric quantile regression models for various quantiles. While we focus on the scenario where the conditional distribution $F(z|x)$ does not depend on $T$, we describe in Section 3 how one can incorporate dependence on the observation time $T$ for situations where $T$ is available. 

Rather than only aiming at the conditional mean, our goal is to target the distribution of slopes at a given level, which  provides insights into the distributional dynamics of the process. Ultimately we target the conditional quantile trajectories of the process $Y$, which describe the probabilistic time dynamics  given a starting point and provide a more comprehensive reflection of the underlying dynamics than the conditional expected trajectories alone.  The assumption that the observed snippets result from realizations of an underlying  stochastic process makes it possible to model  subject-specific variation.

For a concrete example of a process which satisfies (\ref{eq : model}), simple calculations show that smooth monotonic functions with random components are included. For a specific example, consider the model $Y(t) = af(bt + c) + d$ where $f(\cdot)$ is a fixed monotone function, and $(a, b, c, d)$ is a random vector with some joint distribution. Examining the conditional probabilistic behavior of this process at a randomly and independently selected time $T$, and again using the notation $X = Y(T)$, we find
$$ \begin{aligned} F(z|x) &= P(Y'(T) \leq z \bigm| Y(T) = x) \\
&= P(abf'(bT+c) \leq z \bigm| Y(T) = x) \\
&= P\left(abf'\left(f^{-1}\left(\frac{1}{a}Y(T) - \frac{d}{a}\right)\right) \leq z \bigm| Y(T) = x\right) \\
&= P\left(abf'\left(f^{-1}\left(\frac{1}{a}x - \frac{d}{a}\right)\right) \leq z\right), \end{aligned} $$
so that $F(z|x)$ is seen to depend on $T$ through $X=Y(T)$, whence the model in (\ref{eq : model}) is appropriate.\vs

\noindent { \sf 2.2. \quad Evolution of Conditional Distributions and Quantiles}

\no Acquiring an estimate of (\ref{eq : model}) provides insight into the instantaneous probabilistic dynamics of a process. This conditional distribution tells us where subjects generally are headed in the immediate future, based on a certain level, not only in the mean but in distribution. In data applications it is additionally of interest to infer how the process behaves over a longer period of time, beyond the time $T$ where the snippet is recorded.

 Since our goal is to model snippet data in the case where the observed time scale is not informative, taking cross-sectional quantiles from the original snippet data (i.e.~conditioning on $T$) is rendered meaningless in terms of quantifying decline or growth. Instead we focus on a class of quantile models that are based on a given starting level. For this, if full functional trajectories are available, the ideal target would be the cross-sectional distribution of $Y$ for some amount of time $s$ after $Y(T) = x$. Define
\begin{equation}
G_s(y | x) = P(Y(T + s) \leq y \bigm| Y(T) = x),
\label{eq:CSQ}
\end{equation}
as the distribution of $Y(T + s)$ for a given starting value $Y(T)=x$. Taking the $\alpha$-quantile for all $s \in \mathcal{T}$ gives the {\it cross-sectional $\alpha$-quantile trajectory}
\begin{equation}
q_{\alpha, x}(T + s) = G_s^{-1}(\alpha | x).
\label{eq:CSQ2}
\end{equation}
The cross-sectional quantile, when estimable, is a useful tool for data analysis and modeling. 
Selection of the initial value $Y(T) = x$ may vary depending on applications. For example, if information about the baseline status is known, a natural choice would be to model quantile trajectories conditional on starting at baseline. Another interesting method would be to choose individual-specific initial values. Such a model could give practitioners guidance in assessing or ranking individuals which are observed at a certain level and can potentially aid in prediction of an individual's future trajectory. The continuation time $s$ may be interpreted as the time, in the same units as the original data, that has elapsed since the observation time $T$ where the process was recorded at level $Y(T) = x$. In the examples given above, these would be \emph{time since baseline} and \emph{time since observed}, respectively. 

 While the cross-sectional quantile trajectory $q_{\alpha, x}$ is a powerful model, we note that its estimation explicitly depends on observing $Y(T + s).$ In the case of snippet data, the available domain of $s$ where $Y(T+s)$ is observed is very small, making this quantile trajectory an infeasible target. This motivates our proposal to instead  utilize the conditional distribution in (\ref{eq : model})
to assess the long term behavior of the decline process after a random starting time $T$. To this end, we  introduce the notion of an integrated $\alpha$-quantile trajectory, for a given $0 < \alpha < 1$,  and  define \emph{instantaneous $\alpha$-quantiles} for a level $x$ as 
\begin{equation}
\xi_\alpha(x) = F^{-1}(\alpha | x),
\label{eq:InstantQuantile}
\end{equation}
where $F(z|x)$ is defined as in (\ref{eq : model}). 

The function $\xi_\alpha$ describes the $\alpha$-quantile of the conditional distribution of slopes given a level, providing information about the instantaneous rate of decline for the case where trajectories are monotone falling. A simple, yet useful way to utilize and visualize $\xi_\alpha(x)$ is to define a class of trajectories that at all times  follow the $\alpha$-quantile of slope given the current level, thus 
representing a constant quantile of degradation, for example median degradation. 
Accordingly, given $0<\alpha<1$, we define a trajectory $z_{\alpha, x}$ as the solution to an autonomous differential equation
\begin{equation}
\frac{dz_{\alpha, x}(T + s)}{ds} = \xi_\alpha(z_{\alpha, x}(T + s)),
\label{eq:ICQ}
\end{equation}
with initial condition $z_{\alpha, x}(T + 0) = x$. We then refer to the solution function $z_{\alpha, x}(T+\cdot)$ as the  longitudinal $\alpha$-{\it quantile trajectory}   and note that it depends only on $x$ and $\alpha$ and specifically does not depend on the random time $T$. 

Note that $\xi_\alpha$ is the gradient function relating the slope to the level. Defining   longitudinal  quantile trajectories in this way sidesteps $T$, which is unknown.  This approach can be contrasted with  the previous strategy \cp{Abramson1994,Vittinghoff1994} for estimating the conditional mean function $\mu_x(T + s)$, which is the function satisfying 
\begin{equation}
\frac{d\mu_x(T + s)}{ds} = E(Y'(T + s) \bigm| Y(T) = \mu_x(T + s)),
\label{eq:AbrMueller}
\end{equation}
with initial condition $\mu_x(T + 0) = x$. It is straightforward to estimate the $\alpha$-quantile trajectories iteratively, using simple numerical integration techniques, such as Euler's method, to solve the defining autonomous differential equation (\ref{eq:ICQ}). 

 While in general,  longitudinal and cross-sectional quantile trajectories $z_{\alpha,x}$ and $q_{\alpha,x}$ do not coincide, due to basic differences in their definitions, Proposition 1 below demonstrates that $z_{\alpha,x}$ and $q_{\alpha,x}$ will coincide under  some  smoothness and uniqueness assumptions.  Denoting  the space of $k + 1$-times continuously differentiable functions for $k \geq 1$ by $C^{k + 1}$, we assume\vs 

\no
\emph{(A1)} The cross-sectional $\alpha$-quantile  trajectories $q_{\alpha, x}$ satisfy $q_{\alpha, x}\in  C^{k + 1}(\mathcal{T})$ and are monotone in $s$.\\
\emph{(A2)} There exists a function $h \in C^k({\mathcal{J}})$, where $\mathcal{J}$ is the range of $Y$ restricted to $\mathcal{T}$, so that $q_{\alpha, x}$ is the solution to an autonomous differential equation $$\frac{dq_{\alpha,x}(T + s)}{dt} = h(q_{\alpha,x}(T + s)).$$
\emph{(A3)} The cross-sectional quantiles are unique: $G_s(y) = P(Y(T + s) \leq y \bigm| Y(T) = x)$ is strictly monotone in $y$ for all $s \in \mathcal{T}$.\vs

\no Assumptions \emph{(A1)} and \emph{(A2)} ensure that the cross-sectional quantile may be represented by an autonomous differential equation, i.e., that the slope of the conditional cross-sectional quantile only depends on the level. This is a natural assumption in view of (\ref{eq : model}). Assumption \emph{(A3)} ensures that cross-sectional quantiles are well  defined.

\no {\bf Proposition 1.} {\it If the process $Y$ satisfies (A1), (A2), and (A3), then $\alpha$-quantile cross-sectional trajectories and  $\alpha$-quantile  trajectories coincide, i.e., $q_{\alpha, x}(T + s) = z_{\alpha, x}(T + s) \text{ for all } s \in \mathcal{T}.$}

The implication is that under these smoothness and autonomous assumptions we may target $z_{\alpha, x}$ and interpret it as a cross-sectional quantile, which cannot be directly targeted.  This allows us to investigate conditional medians, best and worst case scenarios, and intermediate quantiles, as we will demonstrate in the simulations in Section 5 and the data examples in Section 6. \vs

\bc {\bf \sf 3.\quad ESTIMATION}\sm \ec \rs
\noindent { \sf 3.1. \quad Estimation of Conditional Distributions and Quantiles}

\no The task of estimating $z_{\alpha,x}$ can be decomposed into three steps. First, we estimate the conditional distribution $F(z|x)$. Next we use this estimate to obtain an estimate of the instantaneous conditional quantile function $\xi_\alpha$, according to (\ref{eq:InstantQuantile}). Finally, this estimate of $\xi_\alpha$ is employed as gradient function in an autonomous differential equation as per (\ref{eq:ICQ}), and then this equation is solved numerically. The details are as follows. 

The data in our application is of the form $(X_i, Z_i)$ for $i = 1, \dots, n$, where the snippet data are generated as follows. First, for the $i$th subject, a random mechanism selects the  underlying trajectory  $Y_i$  and an independent random time $T_i$. The $i$-th  subject's  time course is measured in a window $[T_i-\Delta, T_i + \Delta]$ for a small $\Delta >0$,  situated around the random time $T_i$, where the observations in the window are  generated  by a second independent random mechanism as $Y_{ij} = Y_i(T_{ij}) + e_{ij},$ for $T_{ij} \in [T_i - \Delta, T_i + \Delta] \subset \mathcal{T}_0,$ where $e_{ij}$ are independent measurement errors; an illustration of this is in Figure \ref{fig:snippetGen}. For subject $i$ with $n_i$ measurements $Y_{ij}$ at times $T_{ij}$, $j = 1, \dots, n_i$, one may use the empirical estimators $X_i = \frac{1}{n_i}\sum Y_{ij}$ and $Z_i = \hat{\beta}_{1i}$, where $(\hat{\beta}_{0i}, \hat{\beta}_{1i}) = \argmin_{(\beta_{0i}, \beta_{1i})} \sum_{j = 1}^{n_i}(Y_{ij} - \beta_{0i} - \beta_{1i}T_{ij})^2$ for level and slope. Alternatively, if the in-snippet measurements are dense, one can use more sophisticated techniques such as local polynomial regression for estimating level and slope.

The estimation of conditional distribution functions and conditional quantiles has been widely studied in the literature \cp{Hall1999,Li2008,roussas1969, Samant1989, Ferraty2006, Horrigue2011}.  Here we outline several methods by which one can estimate the conditional distributions $F(z|x)$ defined in (\ref{eq : model}),  which is an important auxiliary  target for our method.   We note that if $T$ is known,  one can include it as an additional 
predictor, aiming at   $F(z|x,T)$.

\no \textbf{\emph{Binomial regression.}} Writing  the conditional distribution function as $F(z|x) = E(1(Z \leq z) \bigm| X = x)$ and assuming 
a linear predictor and  a link function $g$, we can model the conditional distribution parametrically as
\begin{equation}
F(z | x) = E(1(Z \leq z) \bigm| X = x) = g^{-1}(\beta_0 + \beta_1x + \beta_2z).
\label{eq:cdfGLM}
\end{equation}
Flexibility can be increased by making use of a generalized additive model
\begin{equation}
F(z | x) = E(1(Z \leq z) \bigm| X = x) = g^{-1}(\alpha_0 + f_1(x) + f_2(z)),
\label{eq:cdfGAM}
\end{equation}
where $f_1$ and $f_2$ are assumed to be smooth with $\int f_1 = \int f_2 = 0$. These parametric methods are simple but require strong assumptions. For instance, the shape of the c.d.f. is determined by the unknown link function $g$. On the positive side, dependence on $T$ and additional covariates can be easily accommodated by including these covariates in the  linear predictor,  making this approach  appealing for some applications.

\no \textbf{\emph{Empirical c.d.f. based on binning.}} A basic approach for nonparametric estimation of a conditional distribution involving two continuous variables is to take a small window around the conditioning variable, selecting the data where the conditioning variable falls into the window, and computing the empirical c.d.f. based on these data. When estimating $P(Z \leq z \bigm | X = x)$, one can simply take a window $\{x \pm h\}$ and calculate the empirical c.d.f. over $Z$ for all subjects satisfying $X \in \{x \pm h\}$, i.e., 
\begin{equation}
\tilde{F}_B(z|x) = \frac{1}{n}\sum_{i = 1}^n 1(Z_i \leq z)1(X_i \in \{x \pm h\}).
\label{eq:cdfB}
\end{equation}
This approach can be extended to include additional covariates but then may be subject to  the curse of dimensionality, depending on the number of covariates. 

\no \textbf{\emph{Kernel smoothing.}} A natural extension of binning is kernel smoothing, analogous to extending a histogram to a smooth density function estimate. This leads to estimators 
\begin{equation}
\tilde{F}_{K}(z|x) = \frac{\sum_{i = 1}^n 1(Z_i \leq z)K_h(x - X_i)}{\sum_{i = 1}^n K_h(x - X_i)},
\label{eq:cdfK}
\end{equation}
where $K_h(u) = h^{-1}K(u/h)$ and $K$ is a kernel function, generally chosen as a smooth and symmetric probability density function.  This estimator is well understood and is a simple extension of the binning method (\ref{eq:cdfB}). Several variants of this estimator have been considered \cp{Hall1999,Li2008}. 

\no \textbf{\emph{Joint kernel smoothing.}} Finally, one might prefer an estimator that is differentiable in $x$ as well as $z$. A smoothed version of (\ref{eq:cdfK}) is obtained by replacing the indicator function with a smooth distribution function $H$:
\begin{equation}
\hat{F}_{JK}(z|x) = \frac{\sum_{i = 1}^n H_{h_H}(z - Z_i)K_{h_K}(x - X_i)}{\sum_{i = 1}^n K_{h_K}(x - X_i)}.
\label{eq:cdfJK}
\end{equation}
Here $K_h$ is as before with bandwidth $h_K$ and $H_{h_H}(u) = H(u/{h_H})$ with possibly different bandwidth $h_H$. For $K$ one commonly uses a probability density function and $H$ is its corresponding cumulative distribution function: $H(u) = \int_{-\infty}^u K(t)dt$. This type of estimator was studied in \citet{roussas1969} and \citet{Samant1989} and has since been extended to include functional predictors in \cite{Ferraty2006} and \citet{Horrigue2011}. As we will require differentiability of our estimate of $F(z|x),$ we will use  estimator (\ref{eq:cdfJK}) in our theoretical results and implementations. 

Given an estimator $\tilde{F}(z|x)$ of ${F}(z|x)$ one can easily construct an estimate of $\xi_\alpha$,
\begin{equation}
\tilde{\xi}_\alpha(x) = \inf\{z:\tilde{F}(z|x) \geq \alpha\},
\label{eq:quantEst}
\end{equation}
which will denote by $\hat{\xi}_\alpha(x)$ if it is based on $\tilde{F}(z|x)=\hat{F}_{JK}(z|x)$  in (\ref{eq:cdfJK}). 

Our  method for estimating $z_{\alpha,x}(T + t)$ involves first using $\tilde{\xi}_\alpha(x)$ in (\ref{eq:quantEst}) as a plug-in estimate for $\xi_\alpha(x)$ in (\ref{eq:ICQ}), and then solving the resulting differential equation using numerical methods. \vs

\noindent { \sf 3.2. \quad Numerical Integration of the Differential Equation}

\no The final estimation step is using an estimate of $\xi_\alpha$ to produce an estimate of $z_{\alpha,x}$. To estimate the solution of (\ref{eq:ICQ}) we may use one of several iterative procedures such as Euler's method or the Runge-Kutta method, along with a uniformly consistent quantile estimate. We first discuss a numerical approximation to the solution $z_{\alpha,x}(T + s)$  and then study its estimation. To simplify notation, in all of the following we assume that $T = 0$, without loss of generality as   $z_{\alpha,x}$ does not depend on $T$.  Then our target becomes $z_{\alpha, x}(s)$ for $s \in \mathcal{T} = (0, \tau]$,  where 
\begin{equation}
\frac{dz_{\alpha, x}(s)}{ds} = \xi_{\alpha}(z_{\alpha, x}(s)),
\label{eq:timeChange}
\end{equation}
with initial condition $z_{\alpha, x}(0) = x$ to guarantee that  the $\alpha$ quantile trajectory  starts at the stipulated level $x$, and again with $\xi_\alpha(x)$ as the $\alpha$-quantile of the distribution of $Y'(T)$ given $Y(T) = x$.

An approximating solution to $z_{\alpha, x}(\cdot)$ is given by $\{s_i, \psi(s_i)\}$, $i = 0, \dots, m,$ for some $m \in \mathbb{N}$, where these quantities are found iteratively using the rule
\begin{equation} \psi(s_0) = x, \quad 
s_{i + 1} = s_i + \delta, \quad 
\psi(s_{i + 1}) = \psi(s_i) + \delta\Phi(\psi(s_i), \delta, \xi_\alpha),\label{eq:numDE} \end{equation}
where $\delta$ is a small time increment. In the case of Euler's method, we have $\Phi(\psi(s_i), \delta, \xi_\alpha) = \xi_\alpha(\psi(s_i))$, while the Runge-Kutta approximation uses $$\Phi(\psi(s_i), \delta, \xi_\alpha) = \frac{1}{6}(k_1 + 2k_2 + 2k_3 + k_4), $$
with
$ k_1 = \xi_\alpha(\psi(s_i)), 
k_2 = \xi_\alpha(\psi(s_i) + \delta k_1 / 2),
k_3 = \xi_\alpha(\psi(s_i) + \delta k_2 / 2), 
k_4 = \xi_\alpha(\psi(s_i) + \delta k_3 / 2)$.

To study the convergence of the numerical solution $\{s_i, \psi(s_i)\}$, consider 
$$ \Delta(s^*, z_{\alpha, x}(s^*), \delta, \xi_\alpha) = \begin{cases} [z_{\alpha, x}(s^* - \delta) - z_{\alpha, x}(s^*)]/\delta &\mbox{if } \delta \neq 0 \\ 
\xi_\alpha(z_{\alpha, x}(s^*)) & \mbox{if } \delta = 0. \end{cases} $$  for a pair $(s^*, z_{\alpha, x}(s^*))$. The  local discretization error at  $(s^*, z_{\alpha, x}(s^*))$ is given by
\be \la{lde}  LDE(s^*, z_{\alpha, x}(s^*), \delta) = \Delta(s^*, z_{\alpha, x}(s^*), \delta, \xi_\alpha) - \Phi(z_{\alpha, x}(s^*), \delta, \xi_\alpha). \ee
For  $-\infty < a_1 < a_2 < \infty,$ the integration procedure defined by $\Phi$ is of order $q$  for an integer 
$q \geq 1$  on $[a_1, a_2]$,  if $LDE(s, z_{\alpha, x}, \delta) = O(\delta^q)$ for all $s \in [a_1, a_2], z_{\alpha, x} \in \mathbb{R}$ and for all $g \in C^q_b(\text{range }z_{\alpha, x}|[a_1, a_2]),$ where $C^q_b[c_1, c_2]$ denotes the set of real functions which are $q$ times continuously differentiable and bounded $q$-th derivative on $[c_1, c_2],$ for $-\infty \leq c_1 < c_2 \leq \infty$. It is well known  that Euler's method achieves order $q=1$, while the Runge-Kutta approximation achieves order $q=4$ \cp{Gragg2006}. 

Of course, the function $\xi_\alpha(x)$ is unknown and must be estimated from the data. We plug in an estimate $\tilde{\xi}_\alpha(x)$ into the right hand side of equation (\ref{eq:ICQ}) and then carry out the numerical integration as described above. This leads to the estimating differential equation
\begin{equation}
\frac{d\tilde{z}_{\alpha, x}(s)}{ds} = \tilde{\xi}_\alpha(\tilde{z}_{\alpha,x}(s)),
\label{eq:estDE}
\end{equation}
for $s \in \mathcal{T} = (0, \tau]$, with initial condition $\tilde{z}_{\alpha, x}(0) = x.$ Using the numerical integration outlined above in (\ref{eq:numDE}) for this differential equation gives the numerical solution $\{s_i, \tilde{\psi}(s_i)\}$. We establish that under regularity conditions, we obtain uniform consistency for $\tilde{\psi}$ as an estimator of $z_{\alpha,x}$ with the corresponding rate depending on the convergence rate of the integration procedure and on  the uniform convergence rate of the estimator $\tilde{\xi}_{\alpha}(x)$. 

\bc {\bf \sf 4.\quad THEORETICAL RESULTS}\sm \ec \rs

\no Our main result is Theorem 2 on uniform consistency of an arbitrary estimate $\tilde{\psi}$ of $z_{\alpha,x}$ given noiseless observations of $Y(T)$ and $Y'(T)$, obtained through numerical integration. Theorem 1 provides consistency for the  estimator $\hat{F}_{JK}(z|x)$ and the associated estimator $\hat{\xi}_{\alpha}$. In particular, we show that $\hat{F}_{JK}(z|x)$ satisfies the assumptions of Theorem 2 and therefore leads to a uniformly consistent estimator of $z_{\alpha,x}.$ We require the following conditions: \vs

\no\emph{(B1)} $\mathcal{J} = \text{range}(Y)|\mathcal{T}_0 = [c_1, c_2]$ and $\mathcal{Z} = \text{range}(Y')|\mathcal{T}_0 = [d_1, d_2]$, where $\mathcal{T}_0$ is the time interval over which measurements are taken. \\
\emph{(B2)} $f_X(x)$, the marginal density of $X$, and $f_{X,Z}(x, z)$, the joint density of $(X, Z)$, satisfy $$ 0 < m_1 \leq \inf_{x \in \mathcal{J}} f_X(x) < \sup_{x \in \mathcal{J}}f_X(x) \leq M_1 < \infty $$
and $$ 0 < m_2 \leq \inf_{x \in \mathcal{J},\, z \in \mathcal{Z}} f_{X,Z}(x,z) < \sup_{x \in \mathcal{J},\, z \in \mathcal{Z}}f_{X,Z}(x, z) \leq M_2 < \infty $$
for some constants $m_1, m_2, M_1, M_2$. \\
\emph{(B3)} With $F^{(i,j)}(x, z) = \frac{\partial^{i+j}F(x, z)}{\partial x^i \partial z^j}$, where $F(x, z)$ is the two-dimensional c.d.f. of $(X, Z)$, $F^{(i + p, j + p)}(x, z)$ exists and is bounded for $(i, j) = (1, 0), (1, 1), (2, 0)$ and $p \geq 2$. Also assume that $f_X(x)$ is $p + 1$ times continuously differentiable.\\
\emph{(B4)} The conditional quantiles are unique, i.e.,  $F(z|x)$ is a strictly monotone function of $z$ in a neighborhood of $\xi_\alpha(x)$. \\
\emph{(B5)} $H'(u) = K(u)$ and $K$ is a symmetric, compactly supported kernel which is $p + 1$-times continuously differentiable, of bounded variation, and satisfies $\int uK(u)du = 0$ and $\int u^2K(u)du < \infty$. \\
\emph{(B6)} The bandwidth $h_K = h_H = h_n$ satisfies (i) $\frac{\log n}{nh_n^2} \rightarrow 0$ as $n \rightarrow \infty$, (ii) $h_n = o((\frac{\log n}{n})^{1/4})$, and (iii) The series $\sum_{n = 1}^\infty \exp\{-\kappa n h_n^4\}$ is convergent for all $\kappa > 0.$ \vs

Assumptions \emph{(B1)} - \emph{(B3)} are standard assumptions regarding the smoothness and boundedness of the distributions when applying smoothing  \cp{Ferraty2006,Hansen2008,Samant1989}. Assumption \emph{(B4)} guarantees that the target quantiles are unique by stipulating that the conditional c.d.f. must not be flat near the $\alpha$ quantile. Finally, assumptions \emph{(B5)} and \emph{(B6)} are typical assumptions for kernel estimators.

{\Theorem Under conditions (B1) - (B6), we have that the estimator $\hat{\xi}_\alpha$, defined at (\ref{eq:quantEst})
 and obtained by inverting $\hat{F}_{JK}(z|x)$ in (\ref{eq:cdfJK})
satisfies
\begin{equation}
\sup_{x \in \mathcal{J}}|\hat{\xi}_\alpha(x) - \xi_\alpha(x)| = O_p\left(h_n + \sqrt{\frac{\log n}{nh_n}}\right)
\label{eq:T1_1}
\end{equation}
and
\begin{equation}
\sup_{x \in \mathcal{J}}|\hat{\xi}'_\alpha(x) - \xi'_\alpha(x)| = o_p(1).
\label{eq:T1_2}
\end{equation}}

For the proof of Theorem 1, one shows first that the estimator $\hat{F}_{JK}(z|x)$ is well-behaved, whence the estimated driving function $\hat{\xi}_{\alpha}$ of the autonomous differential equation in (\ref{eq:estDE}) and its derivative are seen to be consistent. 

For the following main result, some additional  conditions are needed: \vs

\no \emph{(C1)} Assume that one has  a continuously differentiable function  $\tilde{\xi}_\alpha$  satisfying $$ \sup_{x \in \mathcal{J}}|\tilde{\xi}_\alpha(x) - \xi_\alpha(x)| = O_p(\beta_n), \quad \sup_{x \in \mathcal{J}}|\tilde{\xi}'_\alpha(x) - \xi'_\alpha(x)| = o_p(1)$$
for a sequence $\beta_n$ with $\beta_n \rightarrow 0$, $n\beta_n \rightarrow \infty$ as $n \rightarrow \infty$. \\
\emph{(C2)} The function $\Phi(y, \delta, \tilde{\xi}_\alpha)$ defining the numerical integration method (\ref{eq:numDE}) is continuous in its first two arguments on $ G = \{(y, \delta, \tilde{\xi}_\alpha): |z_{\alpha,x}(s) - y| \leq \gamma, 0 \leq s \leq \tau, \delta \leq \delta_0 \}$ 
for some given $\gamma > 0$, $\delta_0 > 0$ and satisfies  
$ |\Phi(y_1, \delta, \tilde{\xi}_\alpha) - \Phi(y_2, \delta, \tilde{\xi}_\alpha)| \leq L|\tilde{\xi}_\alpha(y_1) - \tilde{\xi}_\alpha(y_2)|$ for some $L > 0$ and for all $y_1, y_2, \delta \in G$. \\
\emph{(C3)} The integration method is of order $q$, i.e., the local discretization error satisfies $$ |LDE(s, y, \delta)| = | \Delta(s, y, \delta, \tilde{\xi}_\alpha) - \Phi(y, \delta, \tilde{\xi}_\alpha)| = O(\delta^q)$$
for $s \in \mathcal{T}, \delta \leq \delta_0, y = z_{\alpha,x}(s).$

Assumption \emph{(C1)} is satisfied under the conditions of  Theorem 1 and ensures  consistency of the gradient function estimate which drives  estimation; the condition on the derivative is needed  to control the remainder in the local estimation error. Assumptions \emph{(C2)} - \emph{(C3)} deal with the smoothness and convergence of the numerical integration procedure. We note that both the Euler and Runge-Kutta methods satisfy these requirements.

{\Theorem For an estimator $\tilde{\xi}_\alpha$ of $\xi_\alpha$ which satisfies (C1)  and an integration procedure which satisfies (C2), (C3), we have that the numerical solution $\tilde{\psi}$ of the initial value problem (\ref{eq:estDE}) satisfies
\begin{equation}
\sup_{s \in \mathcal{T}}|\tilde{\psi}(s) - z_{\alpha,x}(s)| = O(\delta^q_n) + O_p(\beta_n),
\label{eq:T2}
\end{equation}
where $\delta_n = |\mathcal{T}|/n$ is the step size used in the integration procedure and $|\mathcal{T}| = \tau$ is the length of the interval $\mathcal{T}$.}

Theorems 1 and 2 imply that under regularity conditions, one can estimate $z_{\alpha,x}$ using the joint kernel described in (\ref{eq:cdfJK}) setting $\tilde{\xi}=\hat{\xi}$ and a well-behaved numerical integration procedure to obtain a uniform convergence rate of $O(\delta_n^q) + O_p(h_n + \sqrt{\frac{\log n}{nh_n}}).$

While in our theoretical analysis we have not considered noise in the measurements, in various applications such noise may be present. This will lead to an errors-in-variables problem that could be of interest for future theoretical work \citep[see][for the conditional quantile case]{Wei2009, Ioannides2009}, while we focus here on the key idea of modeling longitudinal snippets with dynamic systems and the new notion of conditional quantile trajectories. However, we have studied the effects of noise in a simulation in Section 5.1. We also note that if the number of in-snippet measurements increases while the length of the snippets decreases asymptotically, one can use local polynomial smoothing, for example, to obtain consistent estimators of level and slope. Under this scenario, measurement error may be permitted while the convergence rate of $\tilde{\psi}$ will include an additional term.

\bc {\bf \sf 5.\quad FINITE SAMPLE PERFORMANCE}\sm \ec \rs

\no We demonstrate the utility of our conditional quantile methods using a variety of examples in a simulation setting. The first example is a true simulation, where we generate snippets from a process and compare estimates to the true conditional quantile trajectory $z_{\alpha,x}$. Additionally, we consider two real data sets with complete trajectories for which we can visually assess the quality of our estimates. We find that even after reducing the datasets to snippets and ignoring the observation time, we are still able to constuct meaningful trajectories which describe the conditional distribution of outcomes.

\noindent { \sf 5.1. \quad Simulation Study} 

\no To evaluate the performance of our method, we simulate snippet data from an exponential process. This process is generated from sample curves following $ Y(t) = \text{exp}\{-b(t + 1)\} $ where $b$ is a $\mathcal{U}(.3, .5)$ random variable and the time domain is $t \in (0, 10)$.

The sampling procedure to obtain snippets from this random process is as follows. First, a sample of $n$ raw curves $Y_i(t)$ are generated. Next, for each sample trajectory $Y_i$, the sampling time $T_i$ is drawn uniformly and independently over the interval $(0.5, 9.5)$. We take the observation window for $Y_i$ to be $T_i \pm \Delta$, where $\Delta = 0.5$. Over this observation window, we uniformly draw $N_i$ measurements $Y_i(T_{i1}), \dots, Y_i(T_{iN_i})$ where $N_i$ is a random integer in $\{3, 4, 5\}$. In order to investigate the effect of noise, we consider several scenarios: (i) $X_i = Y_i(T_i)$ and $Z_i = Y_i'(T_i)$ are perfectly observed; (ii) $X_i$ and $Z_i$ are not perfectly observed, but $Y_i(T_{ij})$ are sampled noiselessly; (iii) Independent errors $\varepsilon_{ij}$ are  added to $Y_i(T_{ij})$, where $\varepsilon_{ij} \sim N(0, \sigma^2)$.
Surrogates for the levels $X_i$ and slopes $Z_i$ are obtained through  intercepts and slopes of subject-specific linear least squares fits to the data in the window,   as described in Section 3.1.

Our simulation study examines the effect of sample size, choice of $\alpha$, and noise level $\sigma$. For estimation we use the joint kernel conditional c.d.f. estimate as in (\ref{eq:cdfJK}) with $h_K = .01$ and $h_H = .001$ and Gaussian kernels.  We vary the sample size using $n = 300, 1000$ and $5000$. The probability $\alpha$ is varied over $\alpha = \{.10, .25, .5, .75, .90\}$. Finally, the noise levels used are $\sigma = .001, .005,$ and $.01$. All estimated trajectories are conditioned on a starting level of $x = 0.4$.

\begin{figure}[H]
  \centering
  \includegraphics[width = 4in]{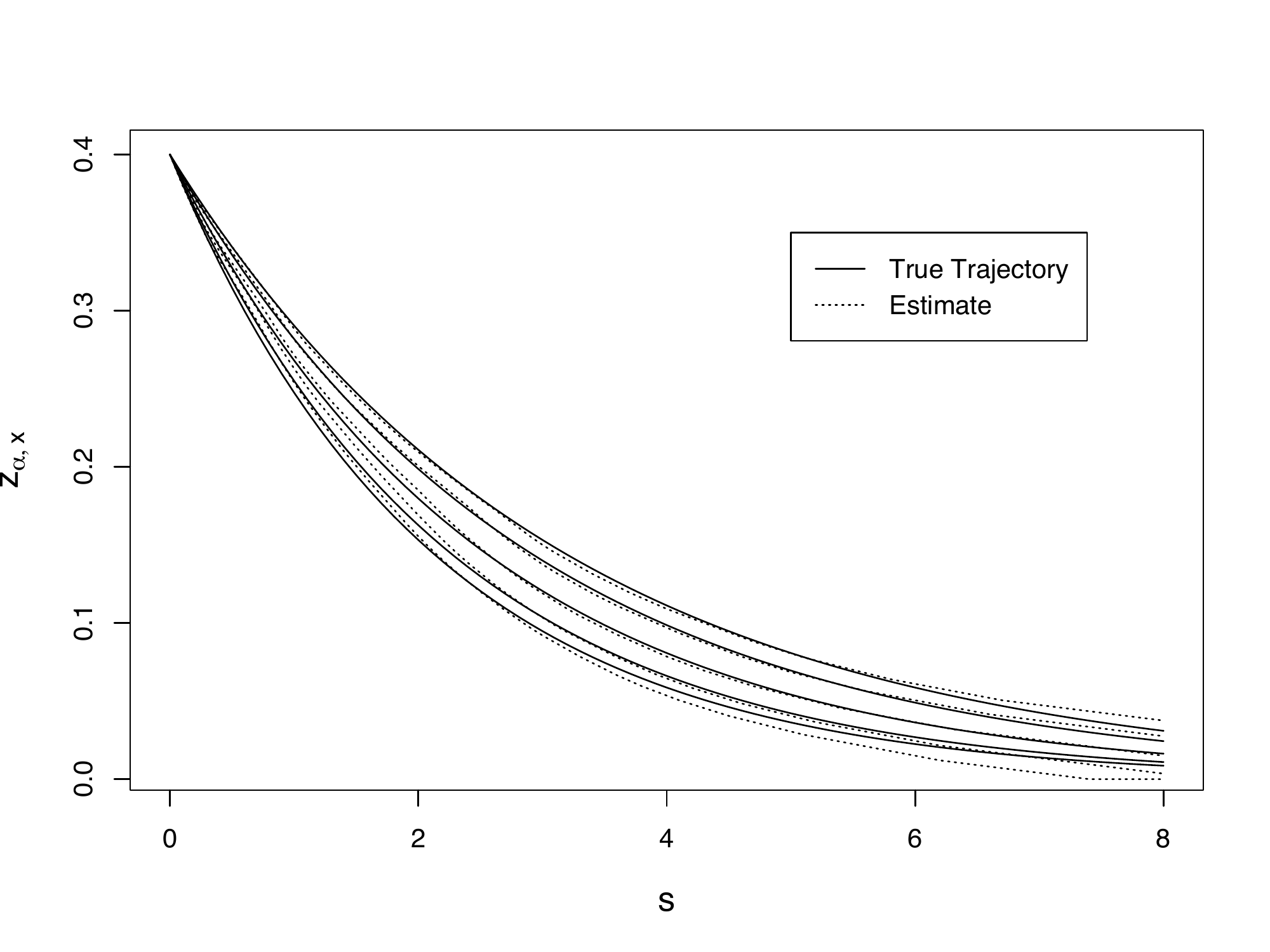}
  \caption{Result of a single simulation with a sample size of $n = 300$, $\alpha \in \{.10, .25, .5, .75, .90\}$, and noise scenario (ii) where longitudinal measurements are taken noiselessly from $Y_i$.}
  \label{fig:simResults}
\end{figure}

As a measure of quality we consider the average integrated squared error (AISE) when repeating the simulations $N = 1000$ times. For our target $z_{\alpha, x}(s)$ for $s \in (0, 8]$ and its estimator, the AISE  is defined as $$ \text{AISE} = \frac{1}{1000}
\sum_{k = 1}^{1000}\int_0^8 (z_{\alpha, x}(s) - \tilde{\psi}^{(k)}(s))^2 ds,$$
where $\tilde{\psi}^{(k)}(s)$ is the functional estimate of $z_{\alpha, x}(s)$ in the $k^{th}$ replication. The time domain in the exponential case is chosen to be (0, 8] as after time $s = 8$ many of the trajectories are flat.  Figure \ref{fig:simResults} demonstrates that our methods perform quite well for $n=300$, especially for the quantiles away from the tails, while the more extreme quantile trajectories are harder to estimate, accounting for the larger spread toward the end of the time domain.

The results for the exponential simulation study  in Table \ref{table:Tapp} reveal that, as expected, the performance of our model declines especially for the more extreme quantiles as the sampling noise increases,  as the accuracy in estimating $(X_i, Z_i)$ is compromised with higher noise levels. This problem can be mitigated with larger sample sizes and potentially more sophisticated techniques for estimating $X_i$ and $Z_i$ when the in-snippet measurements are dense. For further investigation of the noise effect, plots of averaged trajectory estimates are shown in the Supplement in Figure \ref{fig:expAvgd}.

\begin{table}[H]
\centering
\begin{tabular}{c|c||c|c|c|c|c}
 Noise Scenario & Sample Size & $\alpha = .10$ & $\alpha = .25$ & $\alpha = .50$ & $\alpha = .75$ & $\alpha = .90$ \\
 \hline
 \hline
 \multirow{3}{5em}{True $X, Z$} & 300 & 0.275 & 0.078 & 0.091 & 0.085 & 0.132 \\
  & 1000 & 0.295 & 0.045 & 0.047 & 0.068 & 0.127 \\
  & 5000 & 0.299 & 0.037 & 0.036 & 0.063 & 0.125 \\
 \hline
 \multirow{3}{4em}{No noise} & 300 & 0.276 & 0.084 & 0.095 & 0.094 & 0.143 \\
  & 1000 & 0.294 & 0.048 & 0.063 & 0.081 & 0.141 \\
  & 5000 & 0.297 & 0.036 & 0.048 & 0.076 & 0.144 \\
 \hline
 \multirow{3}{4em}{$\sigma = .001$} & 300 & 0.478 & 0.100 & 0.101 & 0.129 & 0.269 \\
  & 1000 & 0.497 & 0.067 & 0.060 & 0.106 & 0.243 \\
  & 5000 & 0.505 & 0.057 & 0.045 & 0.101 & 0.250 \\
 \hline
 \multirow{3}{4em}{$\sigma = .005$} & 300 & 2.220 & 0.654 & 0.127 & 1.042 & 8.471 \\
  & 1000 & 2.189 & 0.635 & 0.052 & 0.827 & 3.935 \\
  & 5000 & 2.209 & 0.634 & 0.031 & 0.801 & 3.603 \\
 \hline
 \multirow{3}{4em}{$\sigma = .01$} & 300 & 6.817 & 2.214 & 0.279 & 6.072 & 43.322 \\
  & 1000 & 6.682 & 2.285 & 0.094 & 4.137 & 22.555 \\
  & 5000 & 6.681 & 2.295 & 0.033 & 3.840 & 18.895 \\
 \hline
\end{tabular}
\caption{AISE scores ($\times 1000$) for various scenarios in the exponential simulation, each with a starting value of $x = 0.4$.}
\label{table:Tapp}
\end{table}

\noindent { \sf 5.2. \quad Simulating Snippets from the Berkeley Growth Data}

\no We also investigate the performance of our methods using growth curves from the Berkeley Growth Study. The Berkeley Growth dataset contains dense growth curves for 39 boys, with measurements spanning ages one to eighteen. To enlarge our sample size, we generate synthetic growth curves by first estimating the mean function as well as the first three eigenfunctions and functional principal component scores for each subject; the first three components accounted for 95\% of the variablity in the original data. We then resample from a nonparametric estimate of the distribution of principal component scores and use these scores to reconstruct a sample of 300 growth curves. 

\begin{figure}[H]
    \centering
  \begin{subfigure}[H]{.4\textwidth}
    \includegraphics[width = \textwidth]{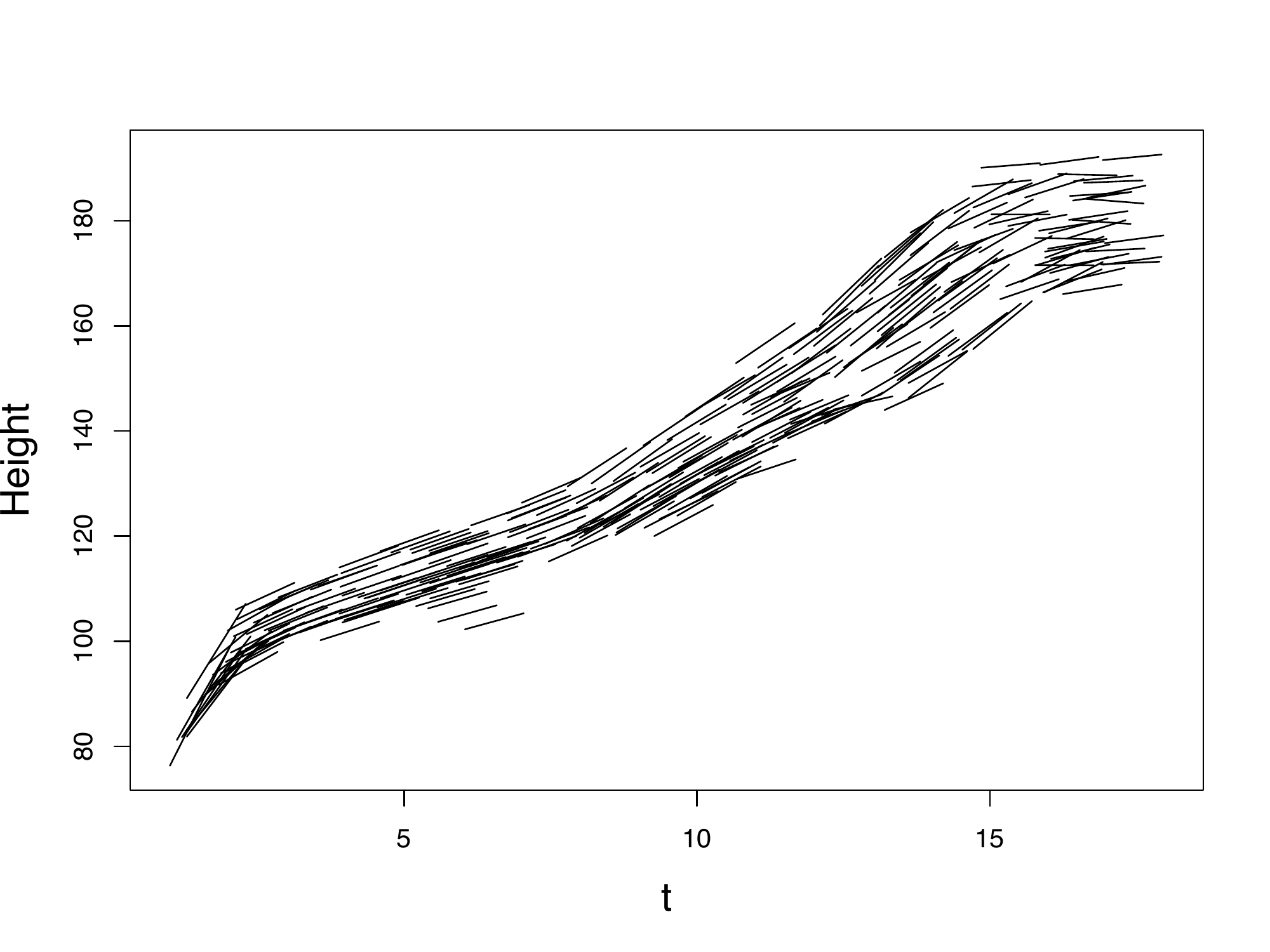}
  \end{subfigure}
  \begin{subfigure}[H]{.4\textwidth}
    \includegraphics[width = \textwidth]{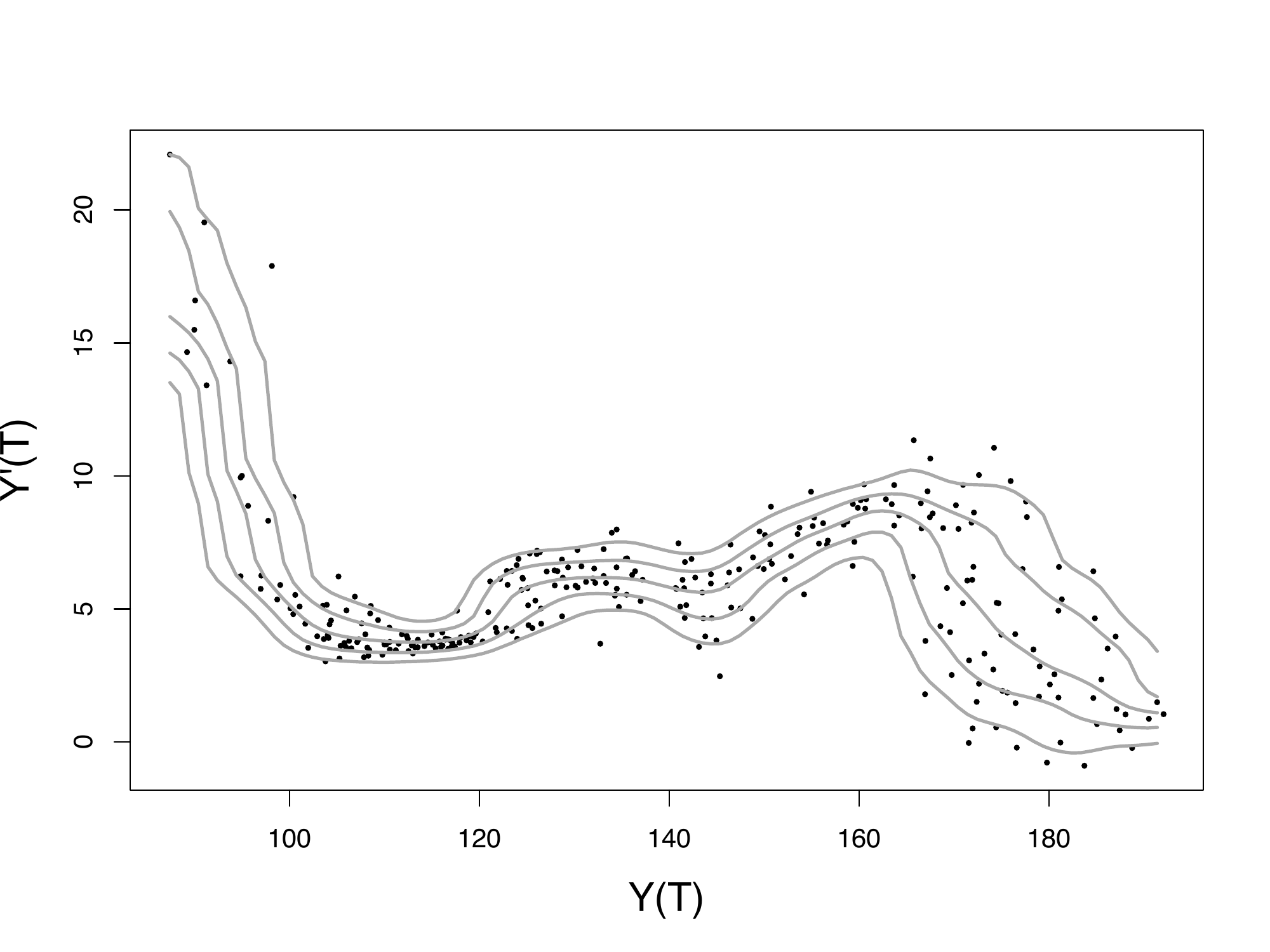}
  \end{subfigure}
  \caption{Artificial snippets created from the Berkeley Growth data (left) and corresponding estimates of $\xi_\alpha(x)$ for $\alpha \in \{.10, .25, .50, .75, .90\}$ (right).}
  \label{fig:berkeleyLevelSlope}
\end{figure}

\begin{figure}[H]
  \begin{center}
    \includegraphics[height = 2.5in]{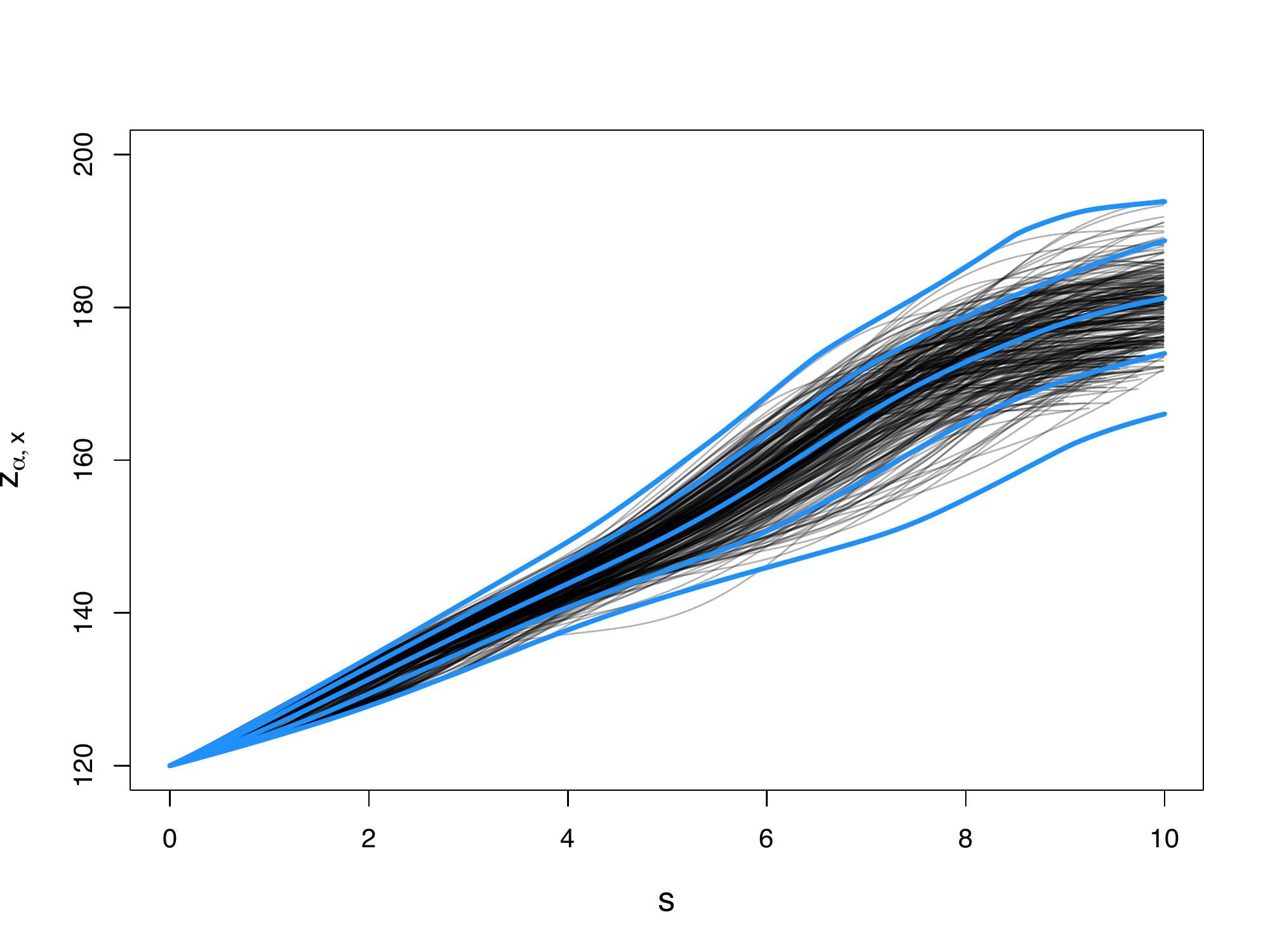}
    \caption{Simulated conditional trajectories  (light grey) sharing the same starting level with the estimated conditional quantile trajectories (blue), which are shown  for quantile levels $\alpha \in \{.10, .25, .50, .75, .90\}$.}
    \label{fig:berkeley}
  \end{center}
\end{figure}

Given  these synthetic growth curves, we create artificial snippets by randomly selecting two measurements, one year apart, for each subject, which are displayed in the left panel of Figure \ref{fig:berkeleyLevelSlope}; the right panel shows a scatterplot of $(X_i, Z_i)$, along with the gradient functions $\hat{\xi}_\alpha(x)$ estimated from the conditional distribution estimate $\hat{F}_{JK}(z|x)$.
Given the level/slope pairs  $(X_i, Z_i)$,  our goal is to estimate the conditional quantile trajectories $z_{\alpha, x}(s)$ for $s \in (0, 10]$, $\alpha \in \{.10, .25, .50, .75, .90\}$, where we condition  on the starting level $x = 120$. We can easily assess how the estimated 
$\alpha$  quantile trajectories  relate to the sample of actual simulated  trajectories by enforcing that each of the functional observations pass through the point $(0, 120)$; see Figure \ref{fig:berkeley}.

Figure \ref{fig:berkeley} demonstrates that the proposed  method reflects nonlinearities in the data quite well, especially as the data that are used in the implementation of the method are very limited. We emphasize that the age at which the subjects entered the study was not used in the estimation; the conditional quantile trajectories were solely estimated from the information in levels and slopes $(X_i, Z_i)$. \vs

\noindent { \sf 5.3. \quad Simulating Snippets from the Six Cities Study of Air Pollution and Growth}

The Six Cities Study of Air Pollution and Growth \citep{Dockery1983}  features  252 subjects, for whom  longitudinal measurements of $\log(\text{FEV1})$, a measure of respiratory function, were taken during childhood over the ages of 6 to 18.
To extract snippets from these data, we randomly select a pair of consecutive measurements for each subject. Our interest is in estimating quantile trajectories, conditional on a starting value of $x = 0.5$. As for the Berkeley growth example, conditional trajectories are found by enforcing that each longitudinal trajectory passes through the point $(0, 0.5)$. 

\begin{figure}[H]
  \begin{center}
  \begin{subfigure}[H]{.49\textwidth}
    \includegraphics[width = \textwidth]{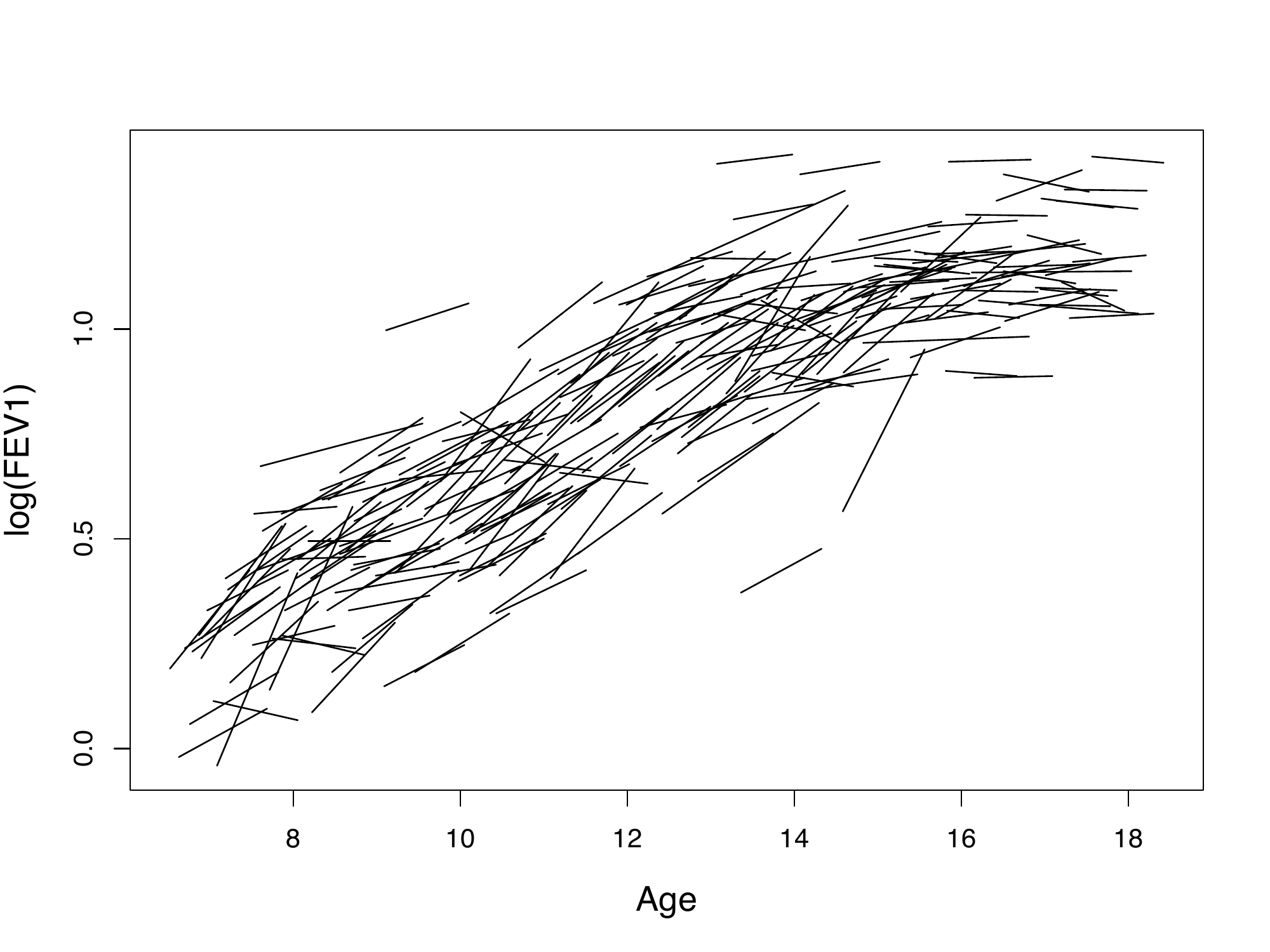}
  \end{subfigure}
  \begin{subfigure}[H]{.49\textwidth}
    \includegraphics[width = \textwidth]{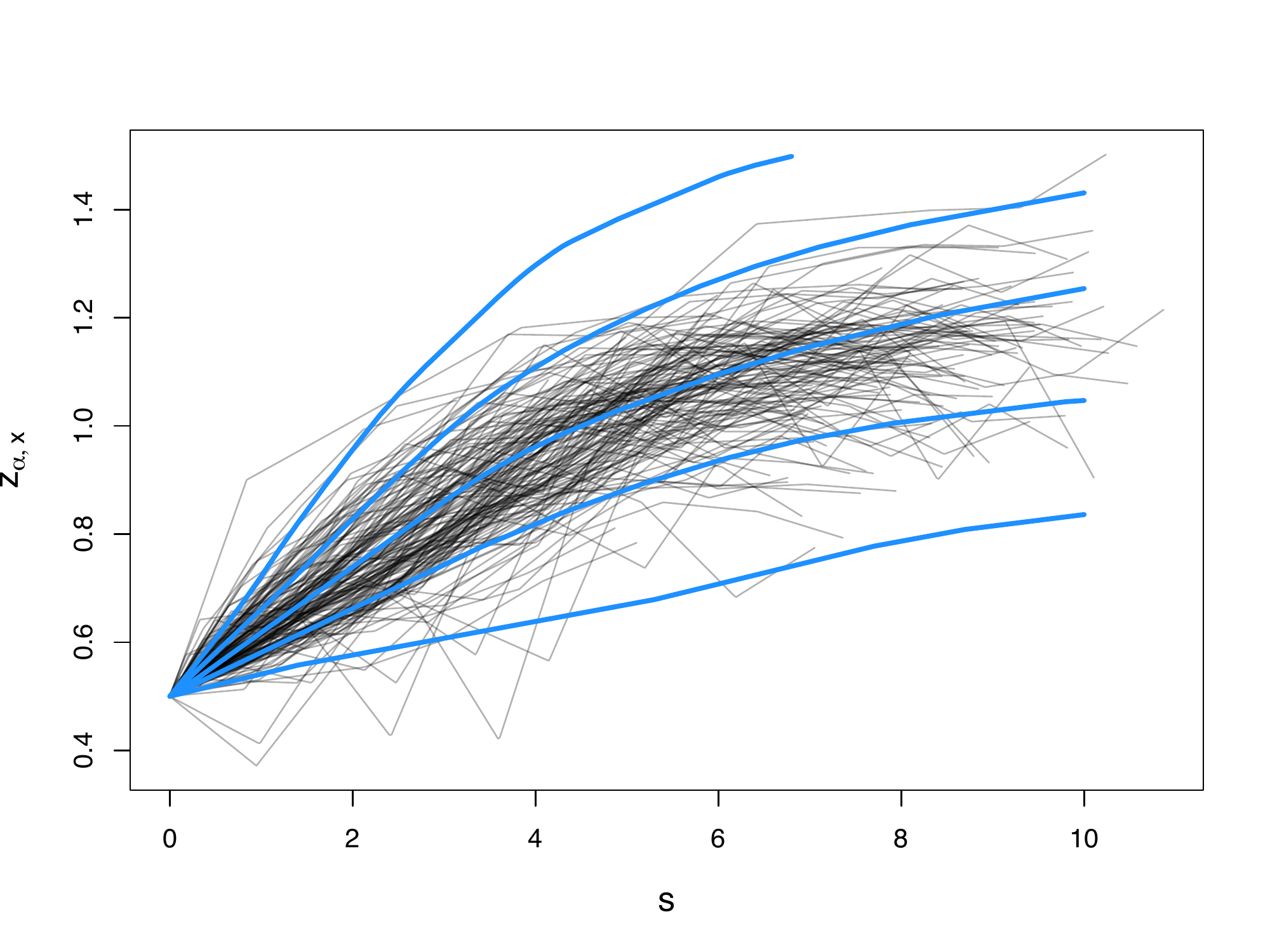}
  \end{subfigure}
    \caption{ FEV1 snippets (left) and conditional quantile trajectories (light grey) extracted from these snippets, where trajectories share the same starting level with the estimated conditional quantile trajectories (blue), shown  for quantile levels $\alpha \in \{.10, .25, .50, .75, .90\}$ (right). }
    \label{fig:fev}
  \end{center}
\end{figure}

\vspace{-1cm}

 Our simulation studies show that the proposed methods work  well for estimating conditional distributions and quantile trajectories for a particular starting level. Indeed, the application to the FEV1 data shows that the methodology described in this paper is suitable for cases where sample sizes are moderate.

\bc {\bf \sf 6.\quad APPLICATION TO ALZHEIMER'S DATA}\sm \ec \rs

It is well established in the Alzheimer's disease (AD) literature that the volume of the hippocampus is decreasing more rapidly for those suffering from Alzheimer's and dementia than it is under normal aging.  The hippocampus is a region in the brain associated with memory,  making hippocampal volume an important biomarker in Alzheimer's diagnosis \citep{Mu2011}. Therefore, one is interested  in modeling the hippocampal volume longitudinally. Here we take the response to be the log hippocampal volume, defined as the log of the sum of the hippocampal lobe volumes (left and right).

An unfortunate aspect of AD is that it can only be diagnosed post-mortem. As such, there is no way of knowing which patients have AD; we can only gain insight via cognitive tests. The subjects in this study have been classified in a clinical evaluation as having normal cognitive function, mild cognitive impairment (MCI), or dementia based on the SENAS (Spanish and English Neuropsychological Assessment Scales)  cognitive test \citep{DMungas2004}. Information regarding the different clinical classifications may be found in \citet{Albert2011,McKhann2011, Sperling2011}.

The right panel of Figure \ref{fig:snippetGen} displays some important features of the dataset of interest, which contains measurements of longitudinal hippocampal volumes for 270 subjects whose ages range from 47 to 96. The snippet characteristics are apparent; with an age range spanning roughly 42 years, the average range of observations per subject averages only 4.2 years. While there seem to be differences between the normal and impaired groups,  it is difficult to describe these differences in terms of mean decline,   as the data form a cloud with no clear patterns. In particular, it is not entirely obvious whether there is an overall decrease over time.   Applying our methods we find  that in fact, there is a clear difference between the groups  and that the rate of hippocampal atrophy is more dramatic than Figure \ref{fig:snippetGen} implies.

\begin{figure}[H]
  \centering
  \begin{subfigure}[H]{.45\textwidth}
    \includegraphics[width = \textwidth]{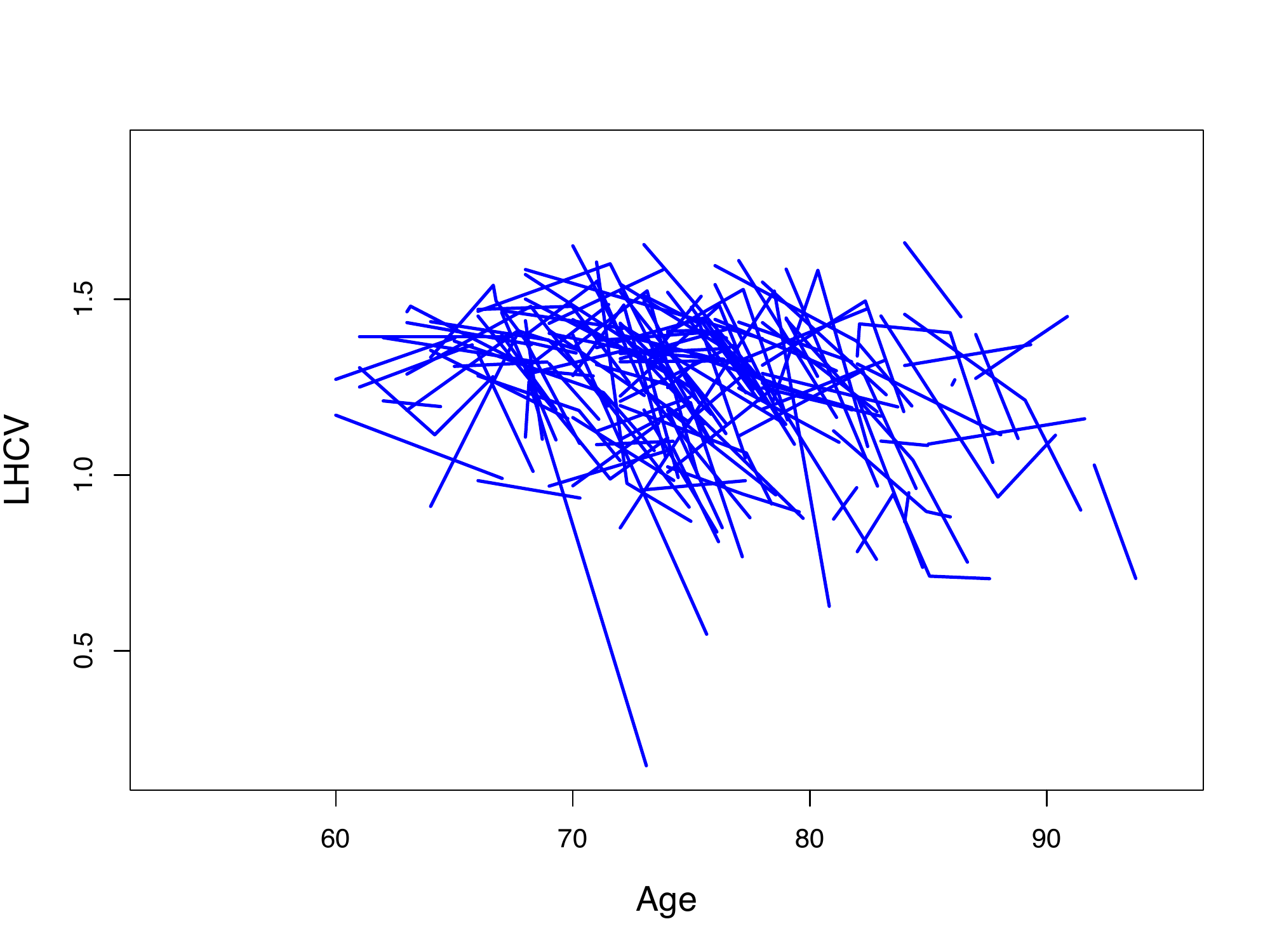}
  \end{subfigure}
  \begin{subfigure}[H]{.45\textwidth}
    \includegraphics[width = \textwidth]{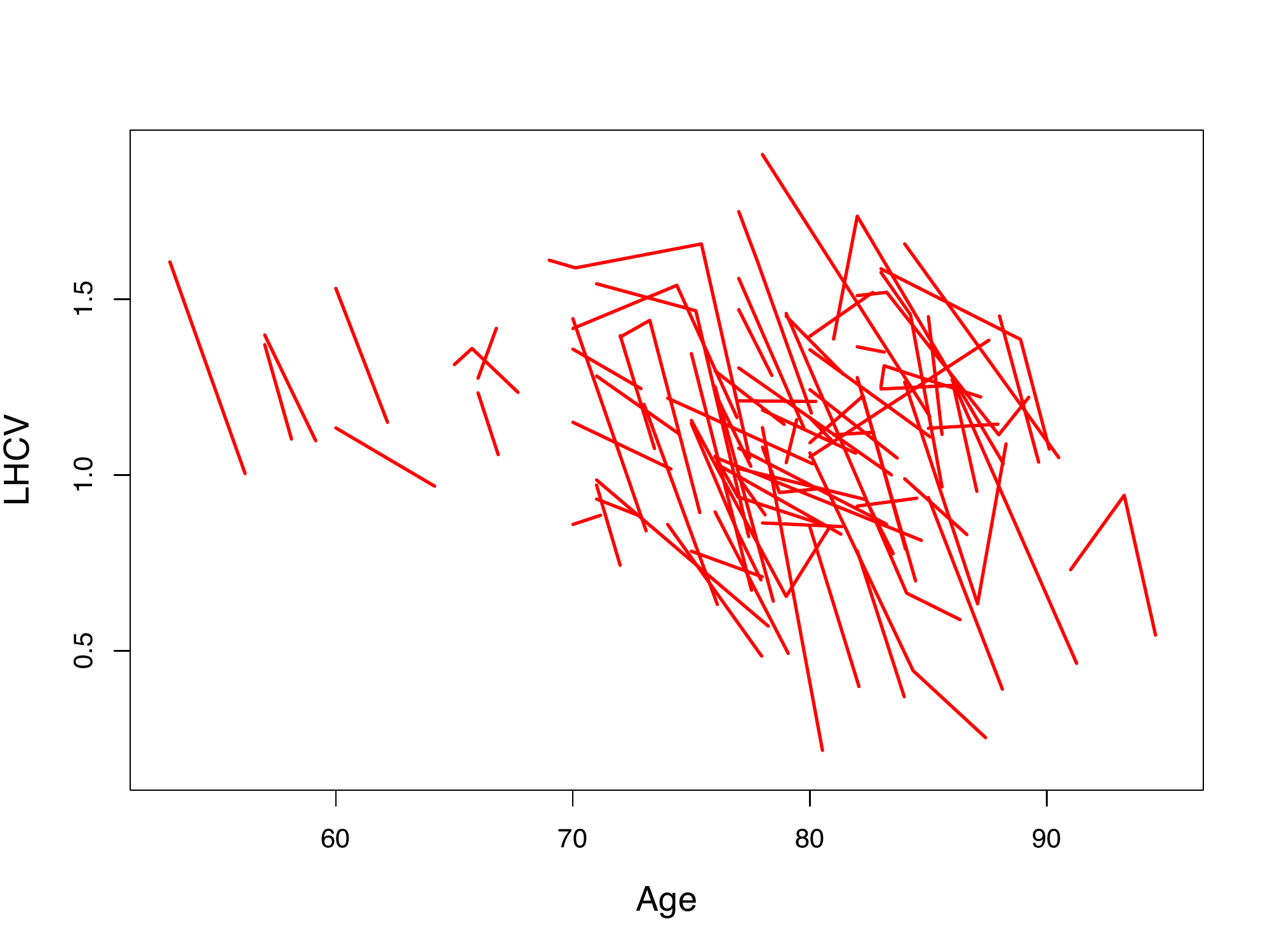}
  \end{subfigure}
  \caption{LHCV longitudinal measurements for normal and demented cognitive groups showing a notable difference between groups.}
  \label{fig:normDemCompare}
\end{figure}

It should be noted that there are several subjects with different cognitive classifications over time. For simplification, we define the impaired subjects as those with at least one MCI or demented classification and the normal group as having no MCI nor demented ratings. The difference between the normal and impaired groups, for example, is pronounced in Figure \ref{fig:normDemCompare} where the groups are plotted separately. While the rates of decline are almost uniformly more severe for the demented group, there seems to be little difference in the mean trend over time with respect to age. Exploratory analysis shows that neither level nor local slope, calculated using a least squares fit on each subjects' measurements, change significantly over chronological age for the demented group. We conclude that chronological age is not informative. 

\begin{figure}[H]
  \begin{center}
    \includegraphics[height = 4in]{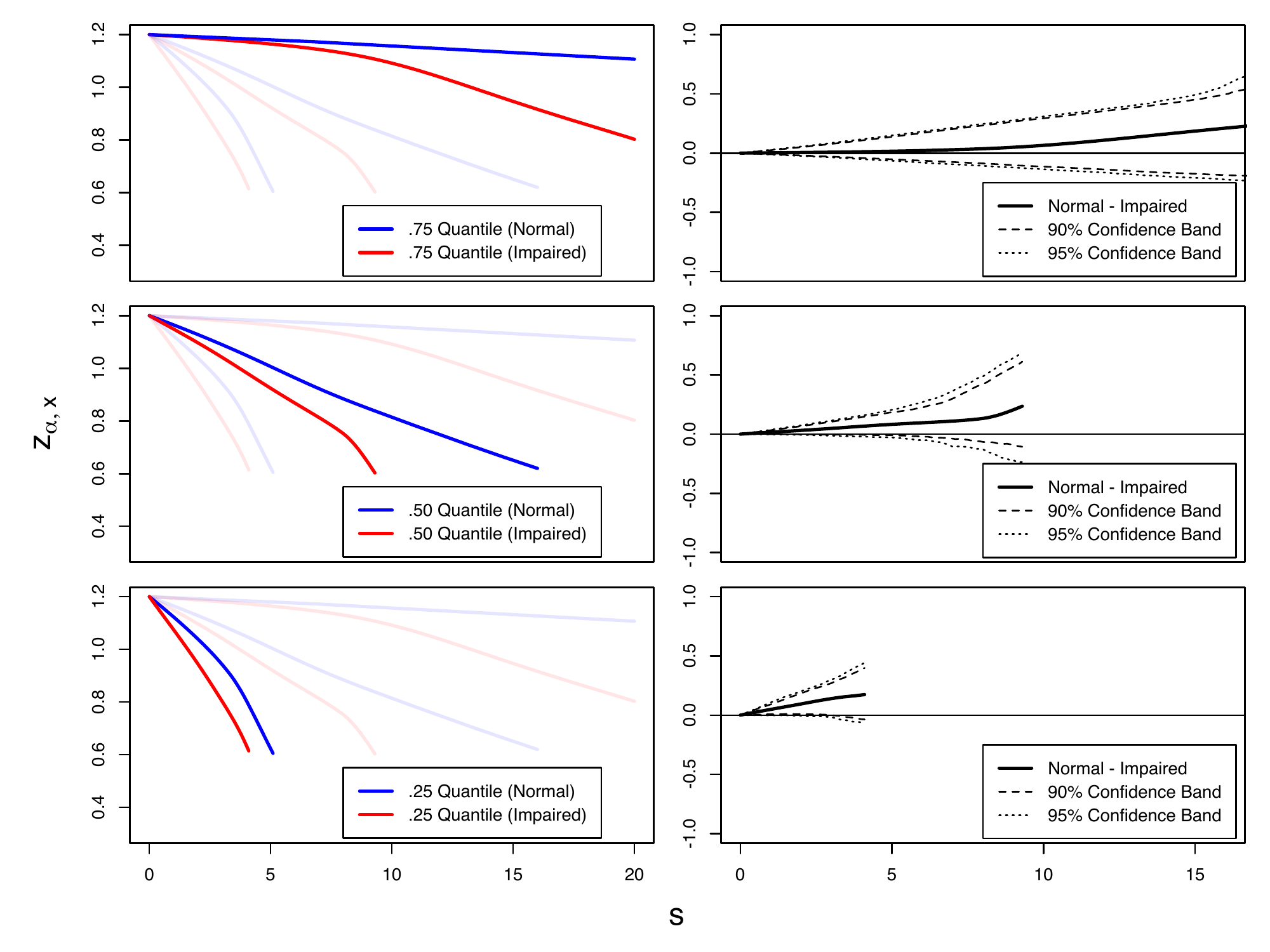}
    \caption{Estimated quantile trajectories conditioned on $x=1.2$ for normal and demented/MCI  groups using the kernel method with bandwidths $h_K = .1$ and $h_H = .01$ and Gaussian kernel (left). From top to bottom, $\alpha$ varies over .75, .50, and .25. The right panels show the difference in trajectories between groups, along with 90\% and 95\% pointwise  bootstrap confidence bands for the difference.}
    \label{fig:allDataZ}
  \end{center}
\end{figure}

The unavailability of a meaningful absolute time measurement is not unique to AD studies. For instance one may be interested in the growth patterns of tumors. Here the time of interest would not be age, but rather the typically unknown time that has passed since the inception of the tumor. In more complete data, one could register the curves based on some landmark features.  With our limited data, however, this is not feasible, and age emerges as an uninformative time scale. 
With little information available to register the data, we therefore need methodology that bypasses age as covariate. Pooling 
demented and MCI subjects, we form a cognitively impaired group and proceed to obtain estimated quantile trajectories  $z_{\alpha, x}$ for varying levels of $\alpha$.   
We use the joint kernel estimator in (\ref{eq:cdfJK}) and Euler's method to obtain quantile trajectories. Our results align with intuition and previous scientific findings. Figure \ref{fig:allDataZ} shows that the decline rate is more severe for the demented/MCI groups than it is for the normal group.

Another interesting result is that hippocampus atrophy is accelerating for lower levels, especially among the demented subjects (see Figure \ref{fig:slopeField} below and Figure \ref{fig:comparison05Plot} in the Supplement). This characteristic implies that the decline of affected subjects is accelerating,  which is a confirmation of what has already been documented in Alzheimer's studies \citep[see][in which the authors caution practitioners from ignoring the nonlinear trend in hippocampal atrophy]{Sabuncu2011}. Figure \ref{fig:allDataZ} shows clear differences in distribution for the two groups, demonstrating future potential trajectories in a data driven way and illuminating  the differences between cognitive groups. 

\begin{figure}[H]
  \begin{center}
    \includegraphics[height = 3in]{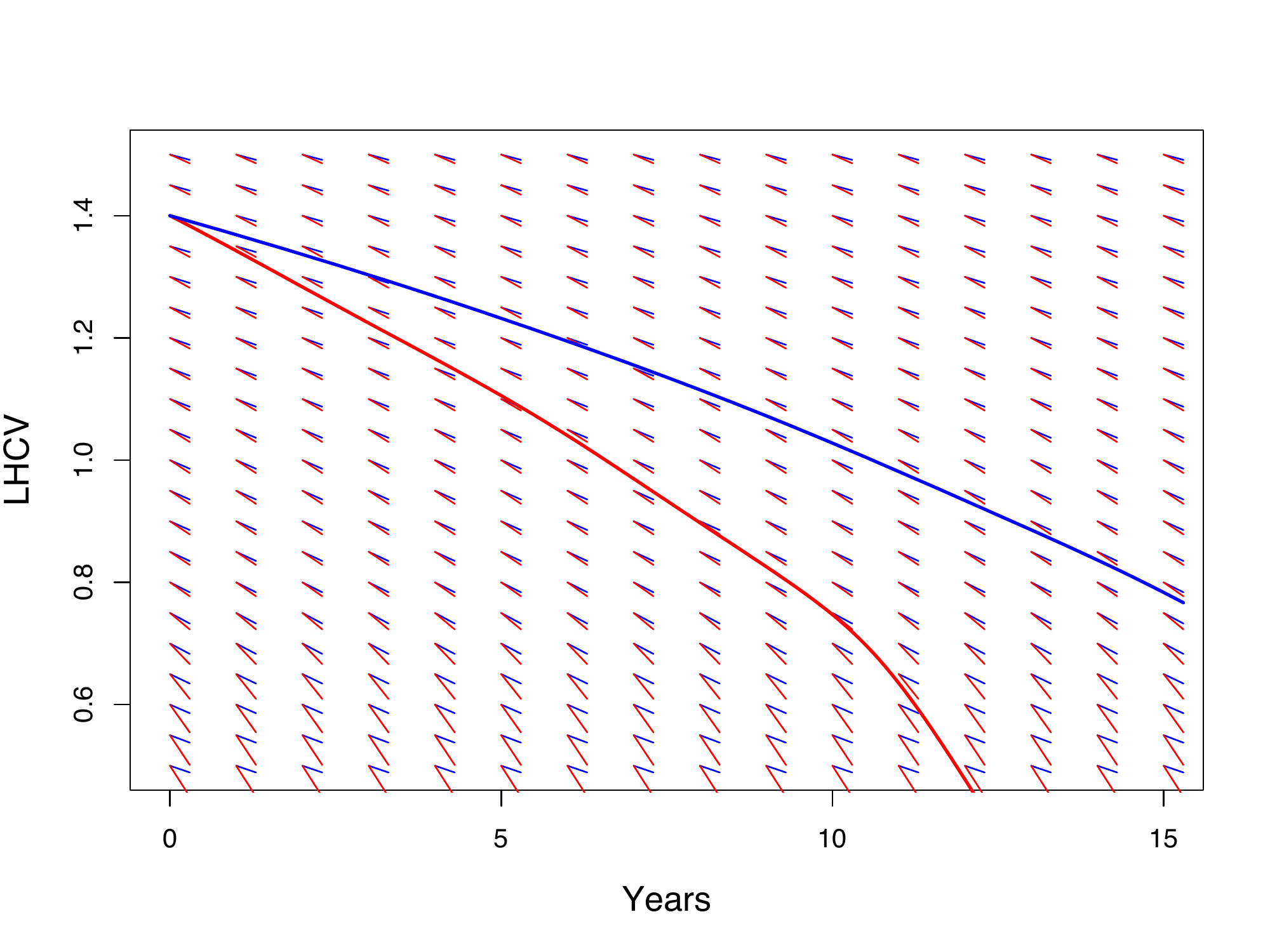}
    \caption{Median slope field for normal (blue) and demented (red) subjects, along with a solution trajectory, conditional on $x = 1.4$}.
    \label{fig:slopeField}
  \end{center}
\end{figure}

\vspace{-.7cm}

\begin{figure}[H]
  \begin{center}
    \includegraphics[height = 4.5in]{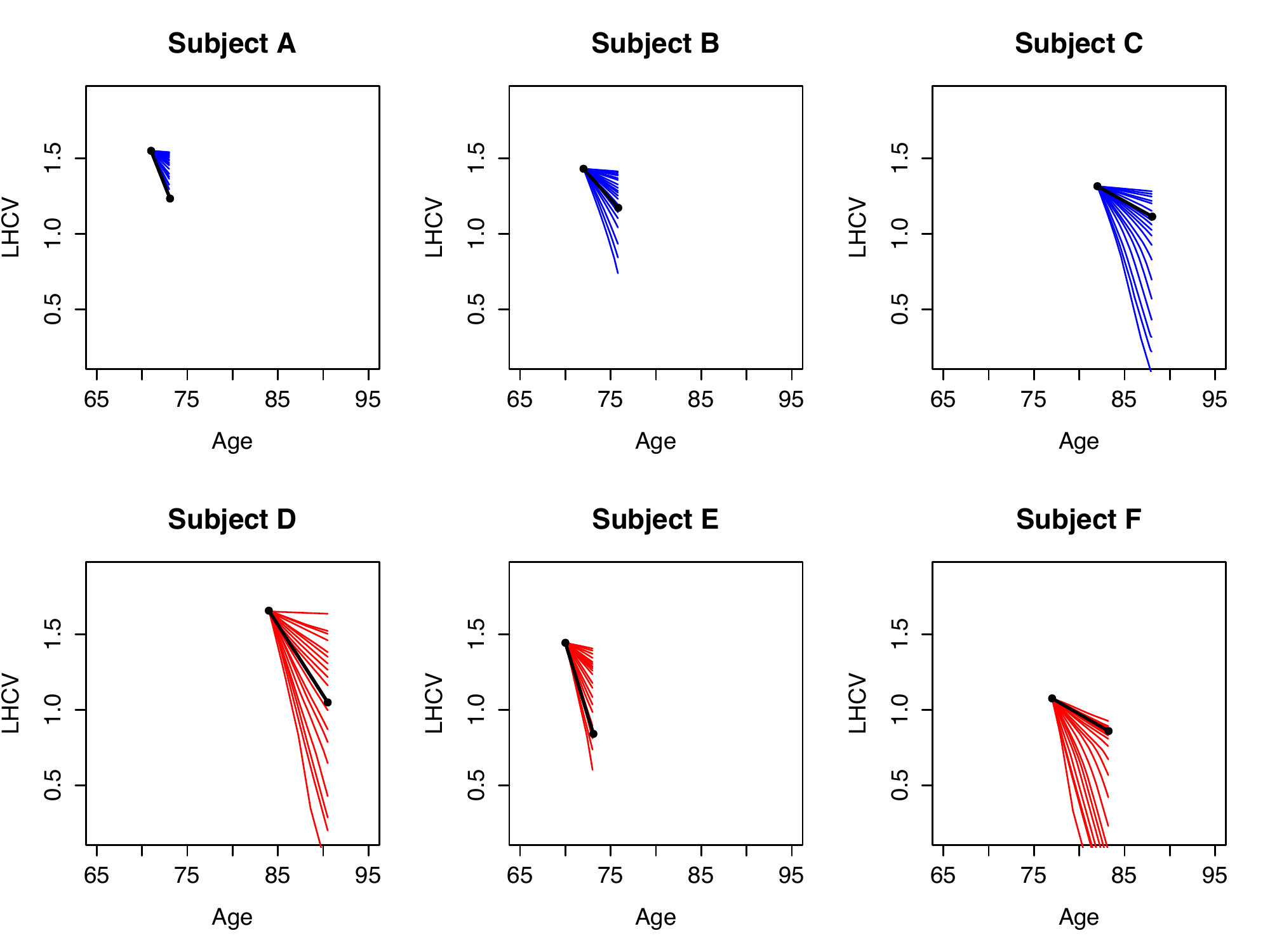}
    \caption{A subset of six individual snippets with corresponding $z_{\alpha, x}$ trajectories originating from the first observation point for each subject. Here, $\alpha \in \{.05, \dots, .95\}.$ The black dots are actual observations for each subject; each of these subjects has two observations. The subjects in the top row are from the normal group while the subjects on the bottom row are from the cognitively impaired group.}
    \label{fig:snippetRanks}
  \end{center}
\end{figure}

It is of great interest to assess individuals by comparing them to the overall sample.  The plots in Figure \ref{fig:allDataZ} represent estimates of population conditional quantiles and can be used to examine the severity of a given subject's trajectory. These comparisons are visualized  for a small sample  of six individuals
(three from each cognitive group)  in Figure \ref{fig:snippetRanks}, where we examine each subject's slope relative to the  estimated conditional quantile trajectories    
 $z_{\alpha, x}$,  as defined in (\ref{eq:timeChange}),  starting from the subject's first observation. This provides   a useful evaluation of  an individual's trajectory, based on pooling information from the entire sample. For example,  normal Subject A is on a very severe trajectory relative  to the normal cognitive group, while Subject F,  though cognitively impaired, is only on a mildly declining trajectory relative to the rest of the cognitively impaired group. 

\vspace{-.7cm}

 We can take our analysis a step further in making ad hoc predictions about a subject's future trajectories. The idea is to estimate conditional $\alpha$ quantile trajectories $z_{\alpha, x}$  for various $\alpha$, starting from the subject's last measurement, thus providing a spectrum of future scenarios for the subject that includes optimistic, median, and pessimistic cases. The practical details are as follows. To ensure that the trajectories align with the subject's snippet, we require that the quantile trajectories are not too far away from  the last observation for the subject. For this, we  first calculate the quantile on which the subject is traveling, and then employ quantile trajectories   $z_{\alpha, x}$ for many values of $\alpha$.  To implement this, we choose  $\alpha = \alpha(s)$ to depend on the time difference between the prediction and the subject's last observation, as follows: If  a subject's snippet corresponds to the level  $\alpha^*$, for estimating the $\alpha$ quantile trajectory we define \begin{equation}\alpha(s) = \begin{cases} \alpha^* + \frac{\alpha - \alpha^*}{S^*}s &\mbox{if } s < S^* \\ 
\alpha & \mbox{if } s \geq S^* \end{cases}, \label{eq:alphaS}\end{equation}
where $S^*$ controls how long the prediction trajectory must adhere to the subject's snippet. In practice, we choose a subject specific $S_i^* = \frac{1}{2}(T_{in_i} - T_{i1})$ to reflect the length of the subject's snippet. We estimate $\alpha^*$ by comparing a subject's slope $Z_i$ to the estimated conditional distribution $\tilde{F}_K(z|y_{i, n_i}),$ where $y_{i, n_i}$ is the subject's last observation. An illustration of $\alpha(s)$ can be found in Figure \ref{fig:alphaPlot} in the Supplement. 

\begin{figure}[H]
  \begin{center}
    \includegraphics[height = 4.5in]{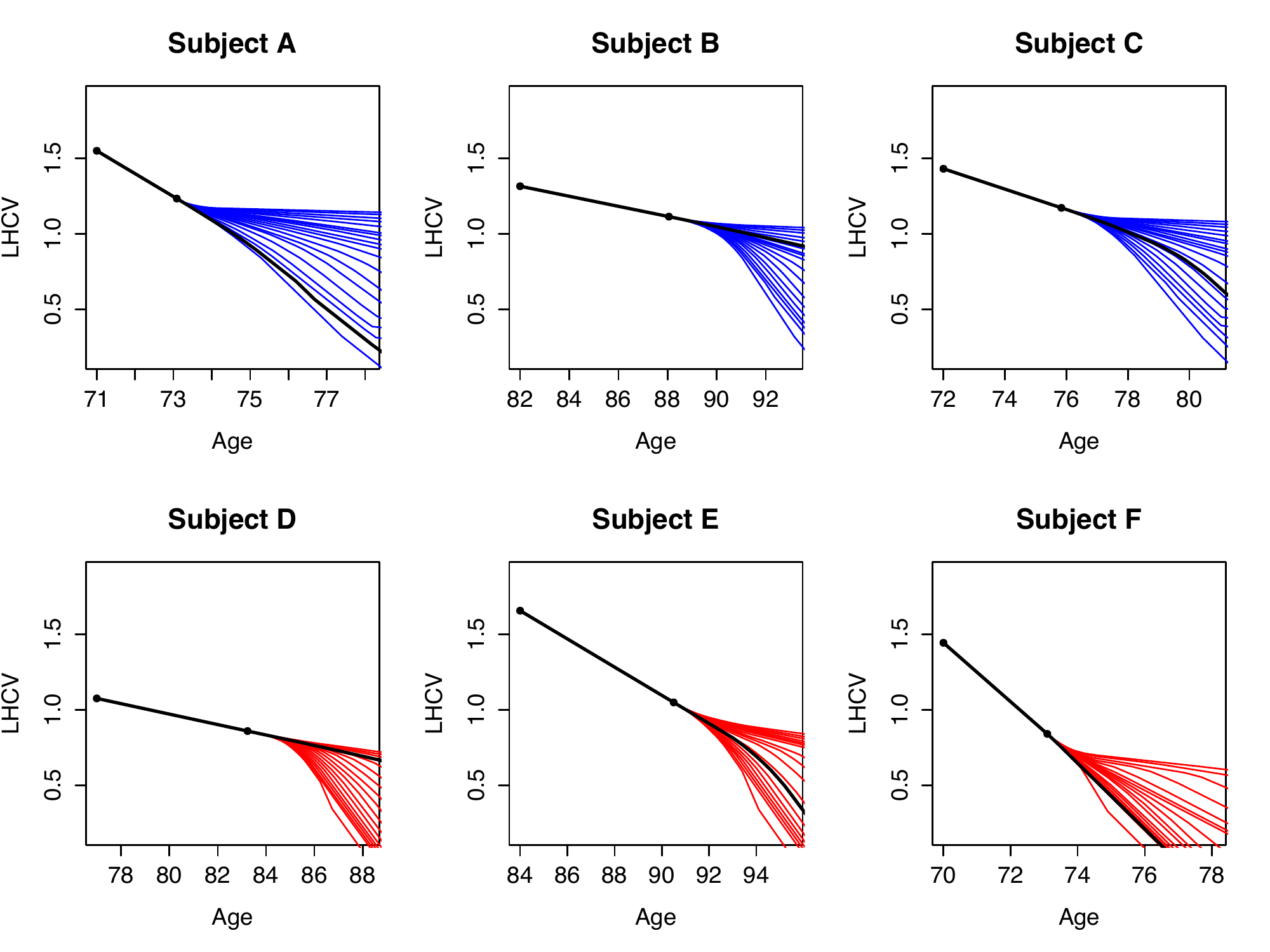}
    \caption{$\alpha$-quantile prediction trajectories where $S^*$ is chosen to be half the length of timespan of the subject's snippet. The black curves that continue the original observed data snippet 
    represent estimated $\alpha^*$ prediction trajectories, where $\alpha^*$ is the estimated quantile of slopes for each subject, conditioned on their last observation. The black dots are the subjects' actual measurements.}
    \label{fig:snippetPrediction}
  \end{center}
\end{figure}

We demonstrate a useful application of the predicted quantile trajectories In Figure \ref{fig:snippetPrediction}. For each subject, we estimate the future trajectory if the subject were to remain at the same quantile $\alpha^*$ throughout; this curve is shown in black. Additionally, we estimate a range of $\alpha$ quantile trajectories under the constraint that early in the prediction, each $\alpha$ must be close to $\alpha^*$. After time $S_i^*$, the $\alpha$ quantile trajectories are unconstrained. 
To demonstrate this prediction method, we use the same six individuals as in Figure \ref{fig:snippetRanks}. This prediction method  is flexible, allows for uncertainty and pools information from the sample while simultaneously enforcing a degree of compliance with a subject's snippet measurements.  

\bc {\bf \sf 7.\quad DISCUSSION}\sm \ec \rs

\no  The problem of longitudinal snippet data is encountered  in accelerated longitudinal medical or social science studies where dense measurements over a long period of time are often not available due to logistical problems, and more generally,  for general functional data when each subject is observed only over a very brief randomly selected time period. For  snippet data one often faces a lack of information about absolute  time, which adds to the challenge in assessing the dynamics of the underlying process.  We distill  the available sparse slope and level information to identify a dynamical system that generates the data and infer information about the random trajectories by introducing the dynamic conditional $\alpha$ quantile trajectories. Our approach relies on the monotonicity of the underlying processes, which makes it possible to adopt the level of the process as a reference as opposed to time. We demonstrate that the  conditional quantile trajectories can be consistently estimated and their estimation,  given an initial value, is straightforward.  

While our methods have been motivated by the challenges posed by an Alzheimer's dataset, we note that the proposed dynamic analysis is applicable to degradation studies or accelerated longitudinal studies when underlying processes are monotone. Instead of examining curves over time, in the monotonic case one can model curves over level. The proposed  methods may also prove useful in and can be easily adapted to the case where (absolute) time is actually available.  For example, in the logistic models described in (\ref{eq:cdfGLM}) and (\ref{eq:cdfGAM}) as well as in the nonparametric settings of (\ref{eq:cdfK}) and (\ref{eq:cdfJK}), age or any other time variable can be easily included in the model as an additional covariate. 
Our methods and especially the instantaneous $\alpha$-quantile and the resulting $\alpha$-quantile trajectory provide useful information about the dynamics of very sparsely observed processes. The proposed estimates perform well and this method will be  a useful tool to recover and predict conditional time-dynamic processes in the presence of limited data.

\newpage
\bc {\bf \sf SUPPLEMENT: PROOFS} \sm \ec \rs
\no\textbf{\emph{Proof of Proposition 1:}}
At $s = 0$ the conditional quantile must be equal to $x,$ i.e., 
$$ q_{\alpha, x}(T + 0) = x. $$
This is because  the conditional quantile is based on a starting value of $Y(T) = x$. Next, since $q_{\alpha, x}$ is smooth, 
$$ \frac{dq_{\alpha, x}(T + 0)}{dt} = \xi_{\alpha}(x),$$
i.e., the slope of the cross-sectional quantile trajectory at $s = 0$ must equal the $\alpha$-quantile of slopes at $s = 0$. From these two facts we have that the slope field that describes the behavior of $q_{\alpha,x}$ at $s = 0$  corresponds to  conditional $\alpha$-quantiles of slopes, since we can condition on any level $x \in \mathcal{J}$. Since the differential equation is autonomous, the starting time is arbitrary so for all $s \in \mathcal{T},$ and for all possible starting levels $x$, we have that the conditional cross-sectional quantile must travel on the longitudinal $\alpha$-quantile trajectory defined in (\ref{eq:ICQ}).

\no\textbf{\emph{Proof of Theorem 1:}}
Equation (\ref{eq:T1_1}) is a direct consequence of theorems in \citet{Ferraty2006} and \citet{Horrigue2011}.

For (\ref{eq:T1_2}), by the Implicit Function Theorem, we have
$$ \xi'_\alpha(x) = -\frac{\partial F(z|x)}{\partial x} \times \left(\frac{\partial F(z|x)}{\partial z} \right)^{-1}\Bigm|_{z = \xi_\alpha(x)}$$
and
$$ \tilde{\xi}'_\alpha(x) = -\frac{\partial \tilde{F}_{JK}(z|x)}{\partial x} \times \left(\frac{\partial \tilde{F}_{JK}(z|x)}{\partial z}\right)^{-1} \Bigm|_{z = \tilde{\xi}_\alpha(x)}.$$

Using the convention that $F^{(i, j)}(x, z) = \frac{\partial F(x,z)}{\partial x^i z^j}$ where $F(x, z)$ is the two-dimensional c.d.f. of ($X, Z$), this leads to

\begin{equation}
\begin{aligned}
\xi_\alpha'(x) &= \frac{f_X'(x)F^{(1, 0)}(x, z) - f_X(x)F^{(2, 0)}(x, z)}{[f_X(x)]^2} \times \frac{f_X(x)}{F^{(1, 1)}(x, z)} \Bigm |_{z = \xi_\alpha(x)} \\
&= \frac{F^{(1,0)}(x,z)f'_X(x) - F^{(2,0)}(x,z)f_X(x)}{f_{X,Z}(x,z)f_X(x)} \Bigm |_{z = \xi_\alpha(x)}
\end{aligned}
\label{eq:T2-xialpha}
\end{equation}
where $f_X(x)$ is the marginal p.d.f. of $X$ and $f_{X, Z}(x, z)$ is the joint p.d.f. of $(X, Z)$.

Applying the same method to the estimator, we have
\begin{equation}
\begin{aligned}
\tilde{\xi}_\alpha'(x) &= \frac{\sum_{i = 1}^n H(\frac{z - Z_i}{h})K(\frac{x - X_i}{h})\sum_{i=1}^n K'(\frac{x - X_i}{h}) - \sum_{i = 1}^n H(\frac{z - Z_i}{h})K'(\frac{x - X_i}{h})\sum_{i=1}^n K(\frac{x - X_i}{h})}{\sum_{i = 1}^n K(\frac{z - Z_i}{h})K(\frac{x - X_i}{h})\sum_{i = 1}^n K(\frac{x - X_i}{h})} \Bigm |_{z = \tilde{\xi}_\alpha(x)} \\
& = \frac{\tilde{F}^{(1, 0)}(x, z)\tilde{f}'_X(x) - \tilde{F}^{(2, 0)}(x, z)\tilde{f}_X(x)}{\tilde{f}_{X, Z}(x, z)\tilde{f}_X(x)} \Bigm |_{z = \tilde{\xi}_\alpha(x)},
\end{aligned}
\label{eq:T2-tildexialpha}
\end{equation}
where 
$$ \begin{aligned}
\tilde{F}^{(2, 0)}(x, z) &\coloneqq \frac{1}{nh^2}\sum_{i = 1}^n H(\frac{z - Z_i}{h})K'(\frac{x - X_i}{h}) \\
\tilde{F}^{(1, 0)}(x, z) &\coloneqq \frac{1}{nh}\sum_{i = 1}^n H(\frac{z - Z_i}{h})K(\frac{x - X_i}{h}) \\
\tilde{f}_{X, Z}(x, z) &\coloneqq \frac{1}{nh^2}\sum_{i = 1}^n K(\frac{x - X_i}{h}) K(\frac{z - Z_i}{h}) \\
\tilde{f}'_X(x) &\coloneqq \frac{1}{nh^2}\sum_{i = 1}^n K'(\frac{x - X_i}{h}) \\
\tilde{f}_X(x) &\coloneqq \frac{1}{nh}\sum_{i = 1}^n K(\frac{x - X_i}{h})
\end{aligned} $$

\no \emph{Lemma:} Under the conditions of Theorem 1, we have
\begin{enumerate}[(i)]
\item $\sup_{x \in \mathcal{J}}|f_X(x) - \tilde{f}_X(x)| = o_p(1)$
\item $\sup_{x \in \mathcal{J}}|f'_X(x) - \tilde{f}'_X(x)| = o_p(1)$
\item $\sup_{x \in \mathcal{J}} \sup_{z \in \mathcal{Z}}|f_{X,Z}(x,z) - \tilde{f}_{X,Z}(x, z)| = o_p(1)$
\item $\sup_{x \in \mathcal{J}} \sup_{z \in \mathcal{Z}}|F^{(1, 0)}(x, z) - \tilde{F}^{(1, 0)}(x, z)| = o_p(1)$
\item $\sup_{x \in \mathcal{J}} \sup_{z \in \mathcal{Z}}|F^{(2, 0)}(x, z) - \tilde{F}^{(2, 0)}(x, z)| = o_p(1)$
\end{enumerate}

\no \emph{Proof of Lemma:} Items (i)-(iii)  follow immediately from \citet{Hansen2008}, while (iv) follows from \citet{Samant1989}. Proving (v) involves only a slight modification of the proof to (iv).

Next, define the numerator in (\ref{eq:T2-xialpha}) as
$$ p(x, z) \coloneqq F^{(1,0)}(x,z)f'_X(x) - F^{(2,0)}(x,z)f_X(x) $$
and similarly for (\ref{eq:T2-tildexialpha}) 
$$ \tilde{p}(x, z) \coloneqq \tilde{F}^{(1, 0)}(x, z)\tilde{f}'_X(x) - \tilde{F}^{(2, 0)}(x, z)\tilde{f}_X(x), $$
as well as the respective denominators 
$$ 
q(x, z) \coloneqq f_{X,Z}(x, z)f_X(x), \quad 
\tilde{q}(x, z) \coloneqq \tilde{f}_{X, Z}(x, z)\tilde{f}_X(x).
$$
The lemma implies that
$$
\begin{aligned}
\sup_{x \in \mathcal{J}}\sup_{z \in \mathcal{Z}}|p(x, z) - \tilde{p}(x, z)| &= o_p(1),\\
\sup_{x \in \mathcal{J}}\sup_{z \in \mathcal{Z}}|q(x, z) - \tilde{q}(x, z)| &= o_p(1).
\end{aligned} $$
Writing 
$$ \tilde{\xi}'_\alpha(x) = \frac{\tilde{p}(x,\tilde{\xi}_\alpha(x)) / q(x, \tilde{\xi}_\alpha(x))}{\tilde{q}(x, \tilde{\xi}_\alpha(x)) / q(x, \tilde{\xi}_\alpha(x))}, $$
we obtain from the lemma
$$ \begin{aligned}
\sup_{x \in \mathcal{J}} \sup_{z \in \mathcal{Z}} \left| \frac{\tilde{q}(x, z)}{q(x, z)} - 1 \right| &= \sup_{x \in \mathcal{J}} \sup_{z \in \mathcal{Z}} \left| \frac{\tilde{q}(x, z) - q(x, z)}{q(x, z)} \right| \\
& \leq \frac{\sup_{x \in \mathcal{J}} \sup_{z \in \mathcal{Z}} |\tilde{q}(x, z) - q(x, z)|}{m_1m_2} = o_p(1).
\end{aligned} $$
Now
$$ \sup_{x \in \mathcal{J}} \sup_{z \in \mathcal{Z}} \left| \frac{\tilde{p}(x, z)}{q(x, z)} - \frac{p(x,z)}{q(x,z)} \right| = \sup_{x \in \mathcal{J}} \sup_{z \in \mathcal{Z}} \left| \frac{\tilde{p}(x, z) - p(x, z)}{q(x, z)} \right| = o_p(1). $$
Since $\frac{\tilde{p}(x, z)}{\tilde{q}(x,z)}|_{z = \tilde{\xi}_\alpha(x)} = \tilde{\xi}'_\alpha(x) \text{ and } \frac{p(x,z)}{q(x,z)}|_{z = \xi_\alpha(x)} = \xi'_\alpha(x),$ we have that
$$ \tilde{\xi}_\alpha'(x) = \xi_\alpha'(x) + o_p(1), $$
uniformly in $x$.

\no\textbf{\emph{Proof of Theorem 2:}}
First we discuss the existence of unique solutions of  the equations
\begin{equation}
\frac{dz_{\alpha,x}(s)}{ds} = \xi_\alpha(z_{\alpha, x}(s)), \hspace{.3in} z_{\alpha,x}(0) = x
\label{eq:T1}
\end{equation}
and
\begin{equation}
\frac{d\tilde{z}_{\alpha, x}(s)}{ds} = \tilde{\xi}_\alpha(\tilde{z}_{\alpha, x}(s)), \hspace{.3in} \tilde{z}_{\alpha, x}(0) = x.
\label{eq:PT2}
\end{equation}
for $s \in \mathcal{T} = [0, T_1]$ for some $0 < T_1 < \infty.$ By the assumptions of the theorem, we have that the functions $\xi_\alpha$ and $\tilde{\xi}_\alpha$ are continuously differentiable over $\mathcal{J}$. It is well known that under these conditions unique solutions exist for both (\ref{eq:T1}) and (\ref{eq:PT2}).

As we are interested in the quantity $\sup_{s \in \mathcal{T}}|\tilde{\psi}(s) - z_{\alpha, x}(s)|$, that is, the supremum difference between the target and its estimate using $\tilde{\xi}_\alpha$ and some integration procedure,
\begin{equation}
\begin{aligned}
\sup_{s \in \mathcal{T}}|\tilde{\psi}(s) - z_{\alpha, x}(s)| &\leq \sup_{s \in \mathcal{T}}|\tilde{\psi}(s) - \tilde{z}_{\alpha,x}(s)| + \sup_{s \in \mathcal{T}}|\tilde{z}_{\alpha,x}(s) - z_{\alpha,x}(s)| \\
&= S_1 + S_2.
\end{aligned}
\label{eq:T3}
\end{equation}

For bounding $S_1$ note that $\tilde{\psi}$ is the numerical approximation of $\tilde{z}_{\alpha, x}$. Since the integration procedure is assumed to be of order $q$, it follows that $S_1$ = $O(\delta_n^q)$ (see the proof of the theorem in \citet{Abramson1994}).

As for $S_2$, consider a sequence of $m$ starting points $s_i = \frac{(i - 1)T_1}{m}$ for $i = 1, \dots, m$ and the following initial value problems with different starting levels:
\begin{equation} \begin{aligned}
\frac{dz_{\alpha,x}(s)}{ds} &= \xi_\alpha(z_{\alpha, x}(s)), \hspace{.3in} z_{\alpha, x}(s_i) = z_{i1}, \hspace{.3in} s \geq s_i, \\
\frac{dz_{\alpha,x}(s)}{ds} &= \xi_\alpha(z_{\alpha, x}(s)), \hspace{.3in} z_{\alpha, x}(s_i) = z_{i2}, \hspace{.3in} s \geq s_i. \end{aligned}
\label{eq:T4}
\end{equation}
Denote the solutions of  these equations as $z_{\alpha,x}(s; z_{i1})$ and $z_{\alpha,x}(s; z_{i2}),$ respectively. These solutions depend continously on the initial conditions $z_{i1}$ and $z_{i2}$ and so there exists a constant $C_1 > 0$ such that
\begin{equation}
|z_{\alpha, x}(s_{i+1}; z_{i1}) - z_{\alpha, x}(s_{i+1}; z_{i2})| \leq C_1|z_{i1} - z_{i2}|
\label{eq:T8}
\end{equation}
for $i = 1, \dots, m$. This controls the difference between two solutions with different initial conditions. Next consider a third initial value problem
\begin{equation}
\frac{d\tilde{z}_{\alpha, x}(s)}{ds} = \tilde{\xi}_\alpha(\tilde{z}_{\alpha, x}(s)), \hspace{.3in} \tilde{z}_{\alpha,x}(s_i) = \tilde{z}_i \hspace{.3in} s \geq s_i,
\label{eq:T5}
\end{equation}
and denote the solution to this equation as $\tilde{z}_{\alpha,x}(s;\tilde{z}_i)$. Next we bound the quantity $$ \sup_{s_i \leq s \leq s_{i+1}}|z_{\alpha, x}(s; \tilde{z}_i) - \tilde{z}_{\alpha, x}(s; \tilde{z}_i)|. $$ By a Taylor expansion, and noting that at the starting time $s_i$, the two functions $z_{\alpha,x}$ and $\tilde{z}_{\alpha,x}$ are the same at $\tilde{z}_i$, 
$$ \begin{aligned}
z_{\alpha,x}(s;\tilde{z}_i) - \tilde{z}_{\alpha,x}(s;\tilde{z}_i) &= z_{\alpha,x}(s_i;\tilde{z}_i) - \tilde{z}_{\alpha,x}(s_i;\tilde{z}_i) \\
&+ (s - s_i)(z'_{\alpha,x}(s_i; \tilde{z}_i) - \tilde{z}'_{\alpha,x}(s_i;\tilde{z}_i)) \\
&+ \frac{(s - s_i)^2}{2}(z''_{\alpha,x}(\zeta; \tilde{z}_i) - \tilde{z}''_{\alpha,x}(\zeta; \tilde{z}_i)), 
\end{aligned} $$
for some $\zeta \in (s_i, s)$, which leads to
\begin{equation}
\sup_{s_i \leq s \leq s_{i + 1}}|z_{\alpha, x}(s; \tilde{z}_i) - \tilde{z}_{\alpha, x}(s; \tilde{z}_i)| \leq \frac{C_2}{m}|\xi_{\alpha}(\tilde{z}_i) - \tilde{\xi}_\alpha(\tilde{z}_i)| + \frac{C_3}{m^2}(C_4 + \sup_{x \in \mathcal{J}}|\xi_\alpha'(x) - \tilde{\xi}_\alpha'(x)|),
\label{eq:T7}
\end{equation}
for $0 \leq i \leq m$ and constants $C_2, C_3, C_4 > 0$.

Combining this  with (\ref{eq:T8}) to get an overall upper bound,  note that for $s \geq s_i$,
\begin{equation}
\tilde{z}_{\alpha,x}(s) = \tilde{z}_{\alpha,x}(s;\tilde{z}_{\alpha,x}(s_i)), \quad 
z_{\alpha,x}(s) = z_{\alpha,x}(s;z_{\alpha,x}(s_i)).
\label{eq:T9}
\end{equation}
Then for $0 \leq i \leq m$, we have
\begin{equation}
\begin{aligned}
\sup_{s_i \leq s \leq_{i + 1}}|z_{\alpha,x}(s) - \tilde{z}_{\alpha,x}(s)| &\leq \sup_{s_i \leq s \leq s_{i+1}}|\tilde{z}_{\alpha,x}(s;\tilde{z}_{\alpha,x}(s_i)) - z_{\alpha,x}(s;\tilde{z}_{\alpha,x}(s_i))| \\ 
&+ \sup_{s_i \leq s \leq s_{i + 1}}|z_{\alpha,x}(s;z_{\alpha,x}(s_i)) - z_{\alpha,x}(s;\tilde{z}_{\alpha,x}(s_i))| \\ 
&\leq O_p\left(\frac{\beta_n}{m}\right) + C_1|z_{\alpha,x}(s_i) - \tilde{z}_{\alpha,x}(s_i)|,
\end{aligned}
\label{eq:T10}
\end{equation}
where the $O_p$ terms are  uniform in $i$. If we choose $m = n$,
$$|\tilde{z}_{\alpha,x}(s_{i + 1}) - z_{\alpha,x}(s_{i+1})| \leq O_p\left(\frac{\beta_n}{n}\right) + C_1|z_{\alpha,x}(s_i) - \tilde{z}_{\alpha,x}(s_i)|.$$
Since the two functions $z_{\alpha,x}$ and $\tilde{z}_{\alpha,x}$ have the same intial condition of $x$ at time $s_0 = 0$,
$$ \sup_{1 \leq i \leq n}|\tilde{z}_{\alpha,x}(s_i) - z_{\alpha,x}(s_i)| = O_p(\beta_n), $$
and therefore
$$ \sup_{s_i \leq s \leq s_{i+1}}|\tilde{z}_{\alpha,x}(s) - z_{\alpha,x}(s)| = O_p(\beta_n), $$
which, since this result is uniform in $i$, implies
$$\sup_{s \in \mathcal{T}}|\tilde{z}_{\alpha,x}(s) - z_{\alpha,x}(s)| = O_p(\beta_n).$$

\bc {\bf \sf SUPPLEMENT: ADDITIONAL MATERIALS} \sm \ec \rs

\begin{figure}[H]
  \begin{center}
    \includegraphics[height = 3.5in]{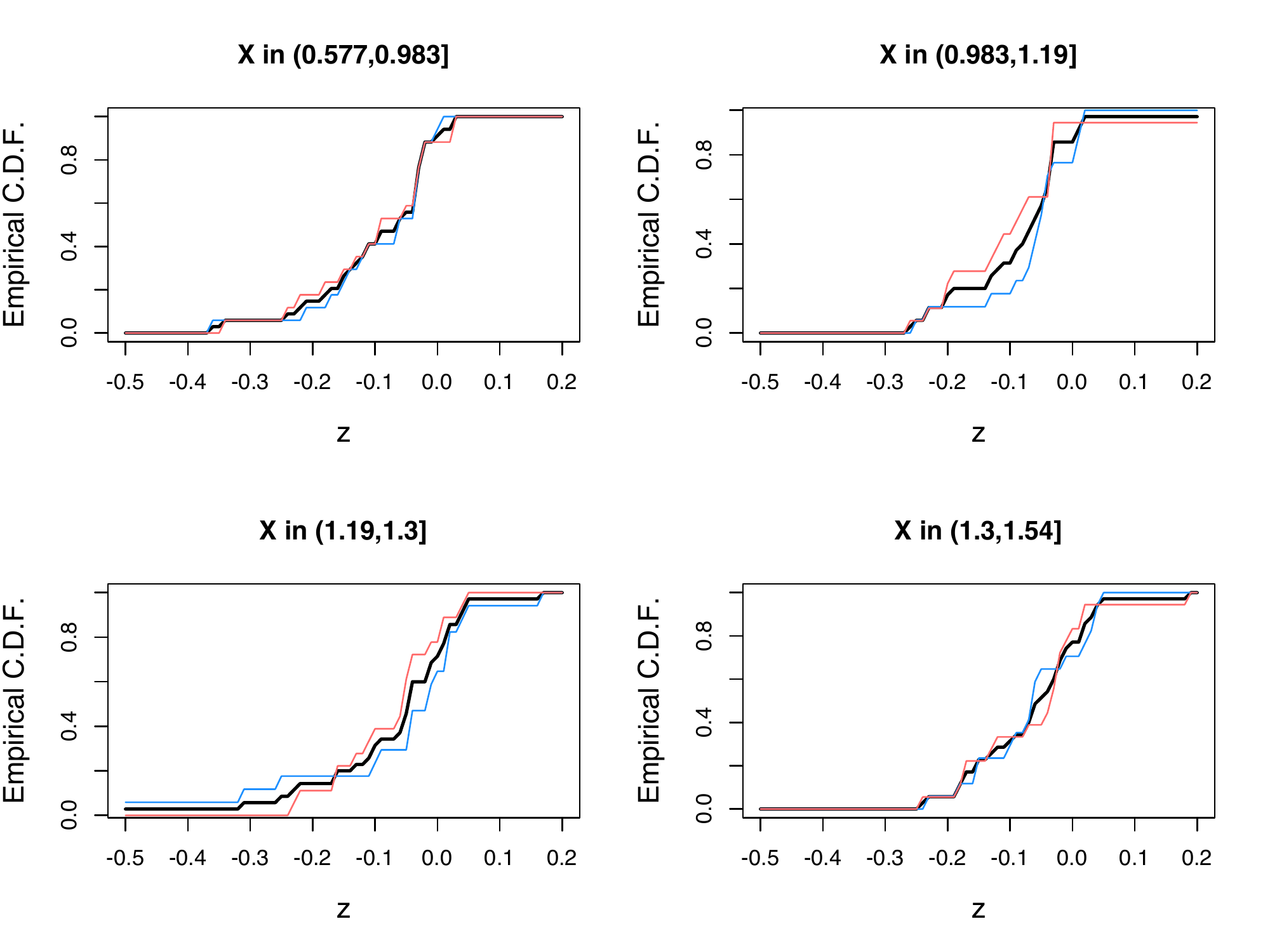}
    \caption{Empirical c.d.f. estimates based on splitting the demented subjects into four level groups, and further splitting into two age groups. Splits are based on quantiles so that the sample sizes in the comparisons are all equal. The overall empirical c.d.f. is shown in black, while the emprical c.d.f.s segmented by age are colored. This demonstrates that the assumption in (\ref{eq : model}) is reasonable. This is also confirmed with Kolmogorov-Smirnov tests, which do not reject for any level segment.}
    \label{fig:empiricalComparison}
  \end{center}
\end{figure}

\begin{figure}[H]
  \centering
  \begin{subfigure}[H]{.4\textwidth}
    \includegraphics[width = \textwidth]{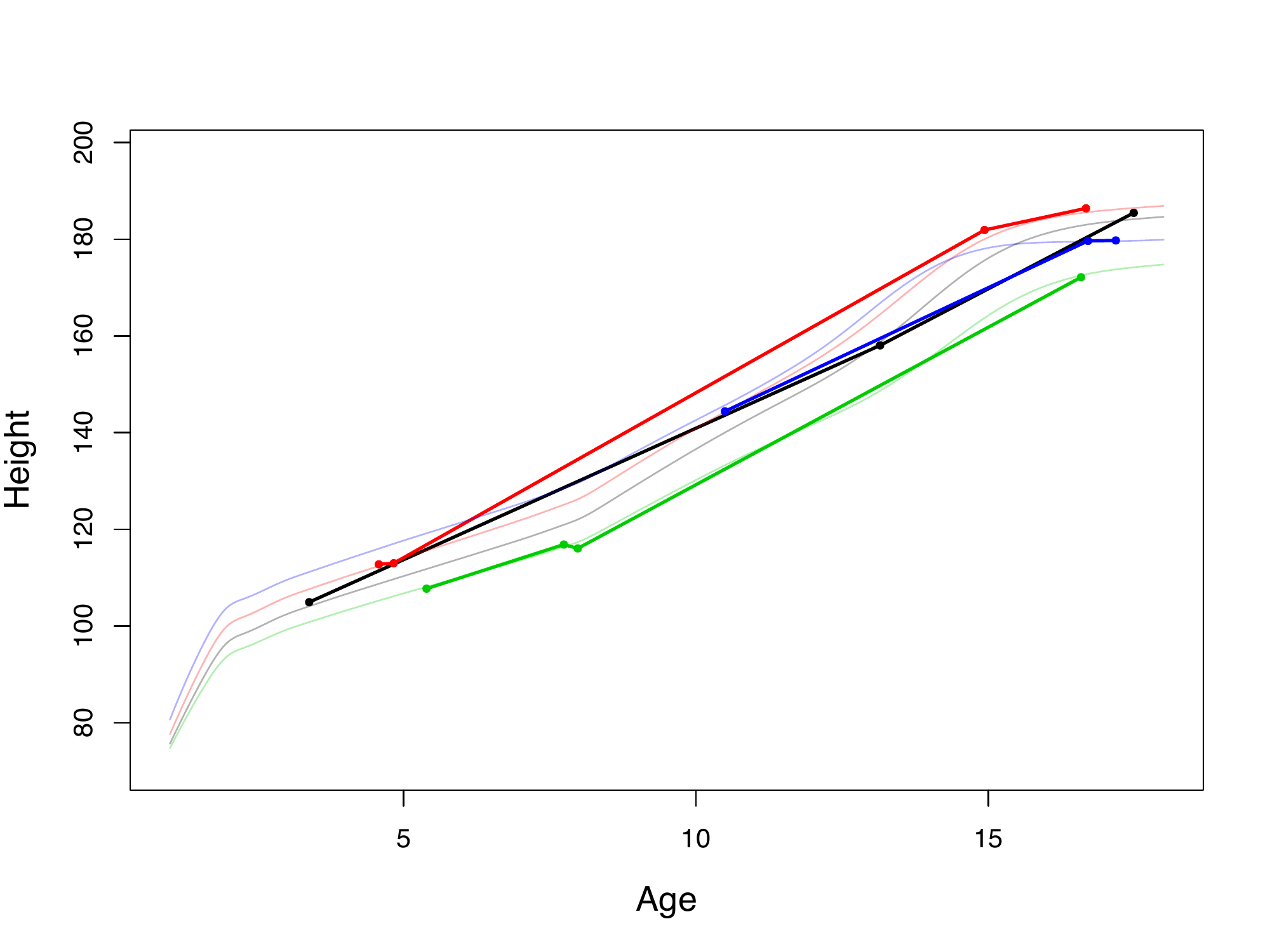}
  \end{subfigure}
  \begin{subfigure}[H]{.4\textwidth}
    \includegraphics[width = \textwidth]{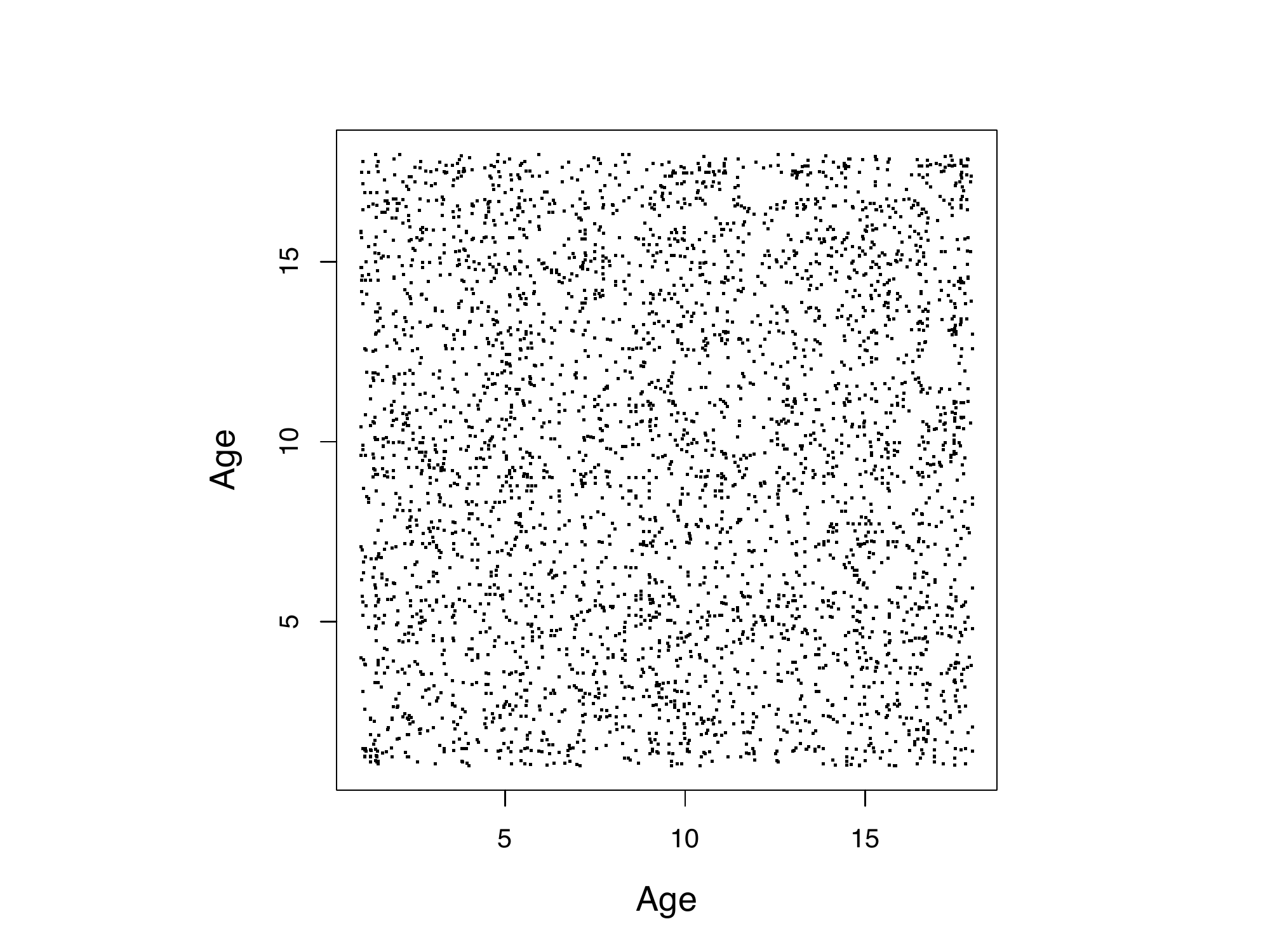}
  \end{subfigure}
  \begin{subfigure}[H]{.4\textwidth}
    \includegraphics[width = \textwidth]{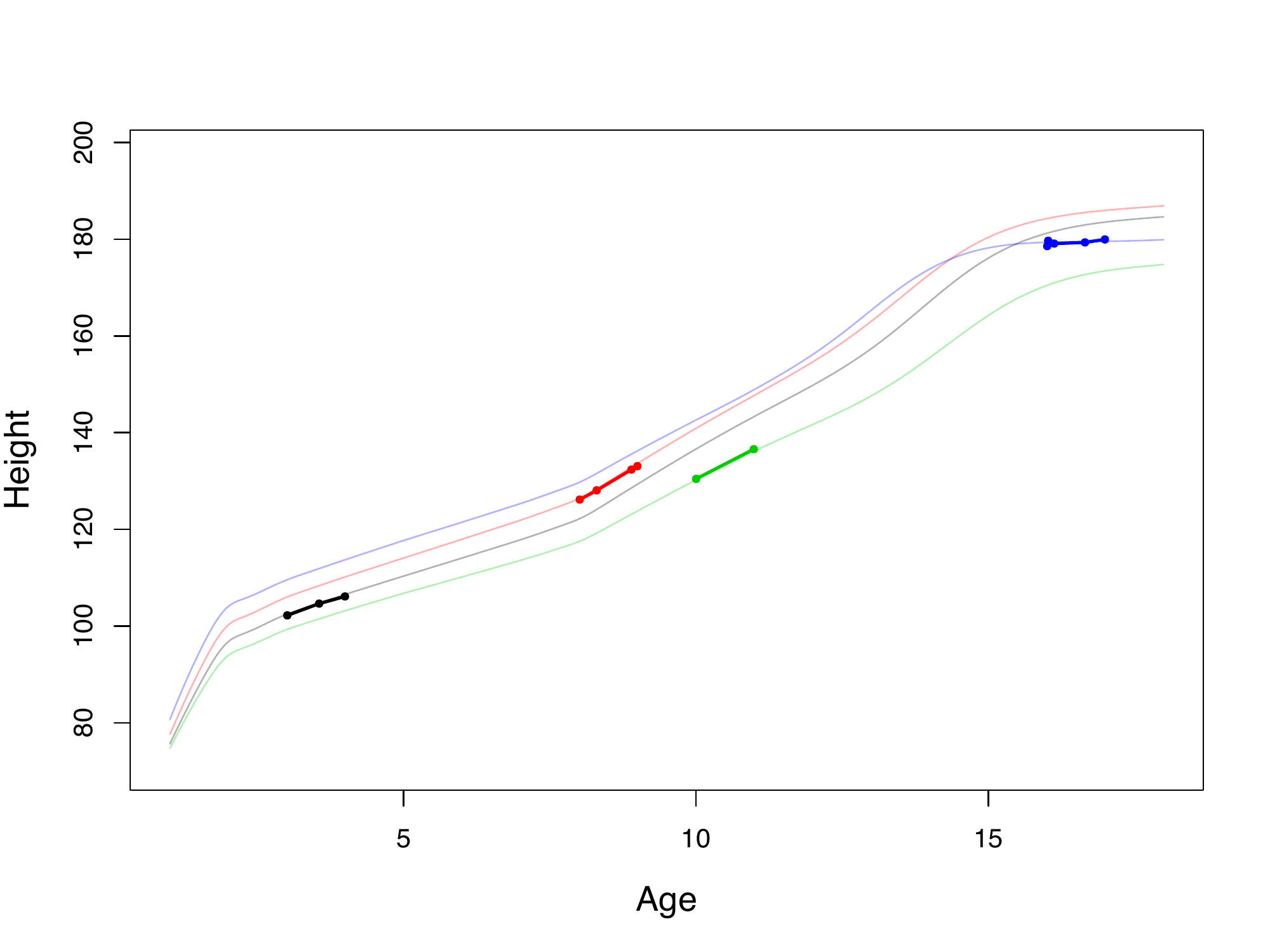}
  \end{subfigure}
  \begin{subfigure}[H]{.4\textwidth}
    \includegraphics[width = \textwidth]{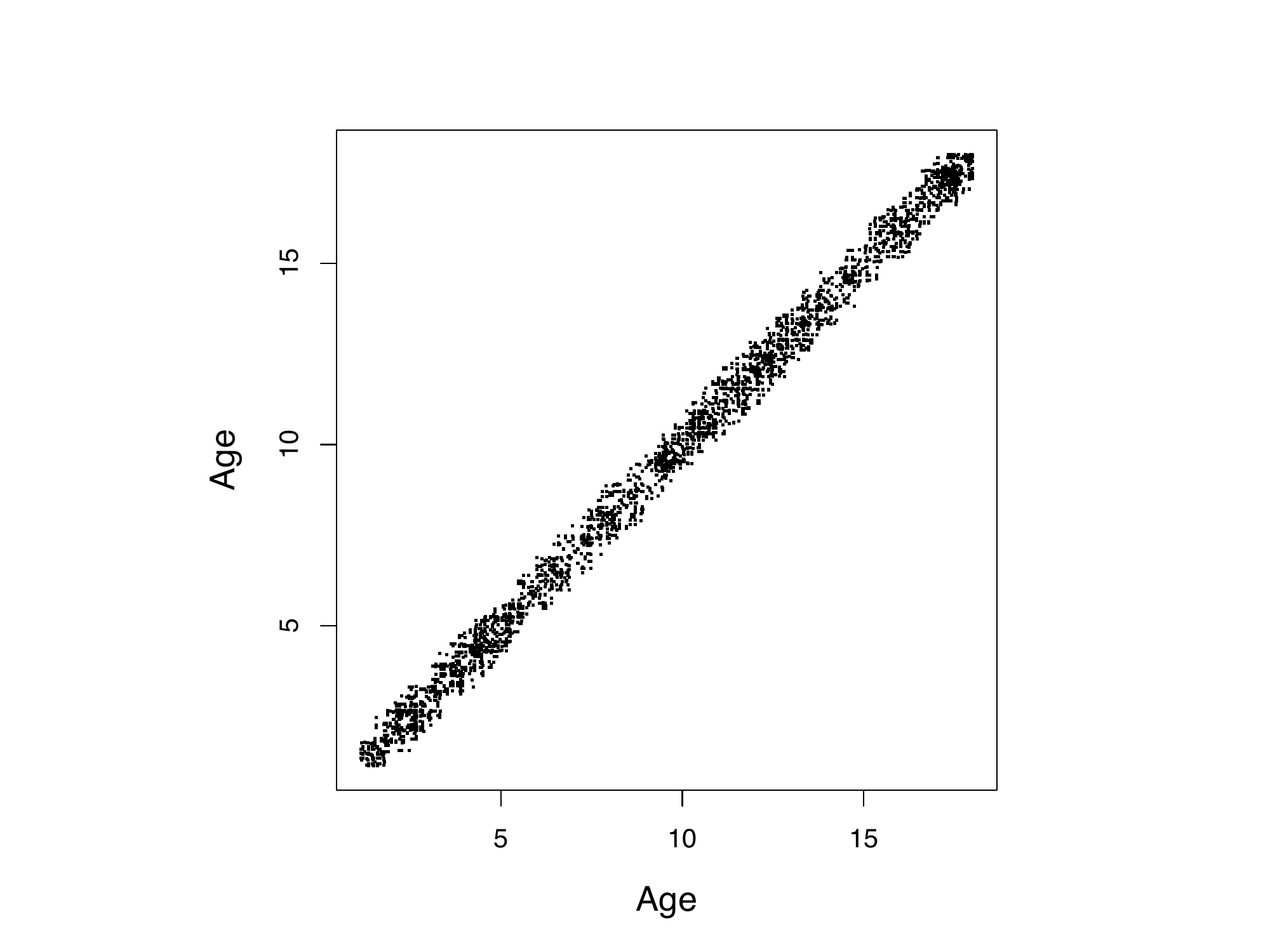}
  \end{subfigure}
  \caption{A comparison between sparse longitudinal data (top) and snippet data (bottom) showing a few longitudinal observations (left) and full sample design plots (right). Data were simulated based on the first three functional principal components of Berkeley Growth curves. Design plots are based on 300 sparse or snippet trajectories, respectively.}
  \label{fig:sparseVSnippet}
\end{figure}

\begin{figure}[H]
  \begin{center}
  \begin{subfigure}[H]{.45\textwidth}
    \includegraphics[width = \textwidth]{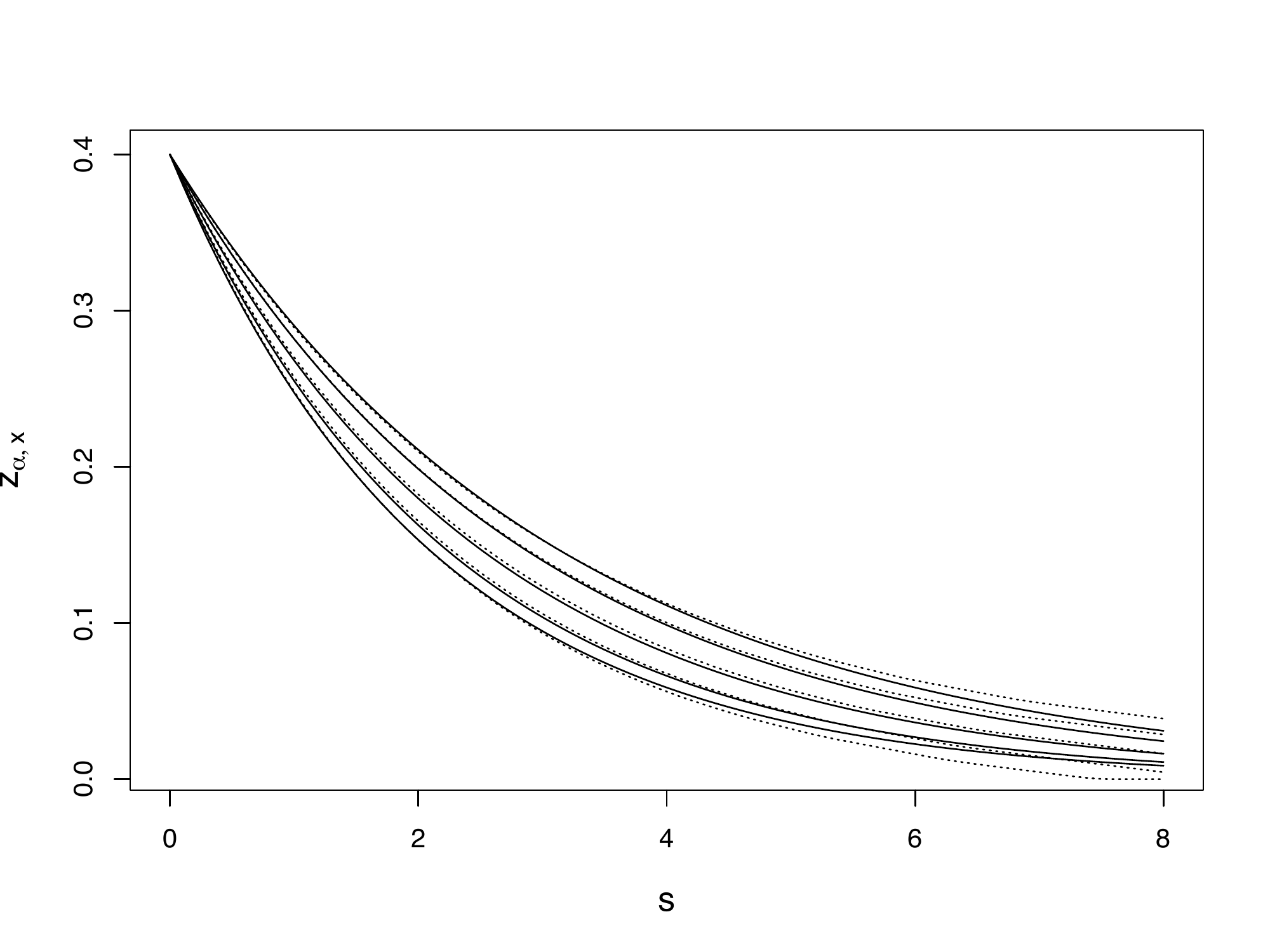}
  \end{subfigure}
  \begin{subfigure}[H]{.45\textwidth}
    \includegraphics[width = \textwidth]{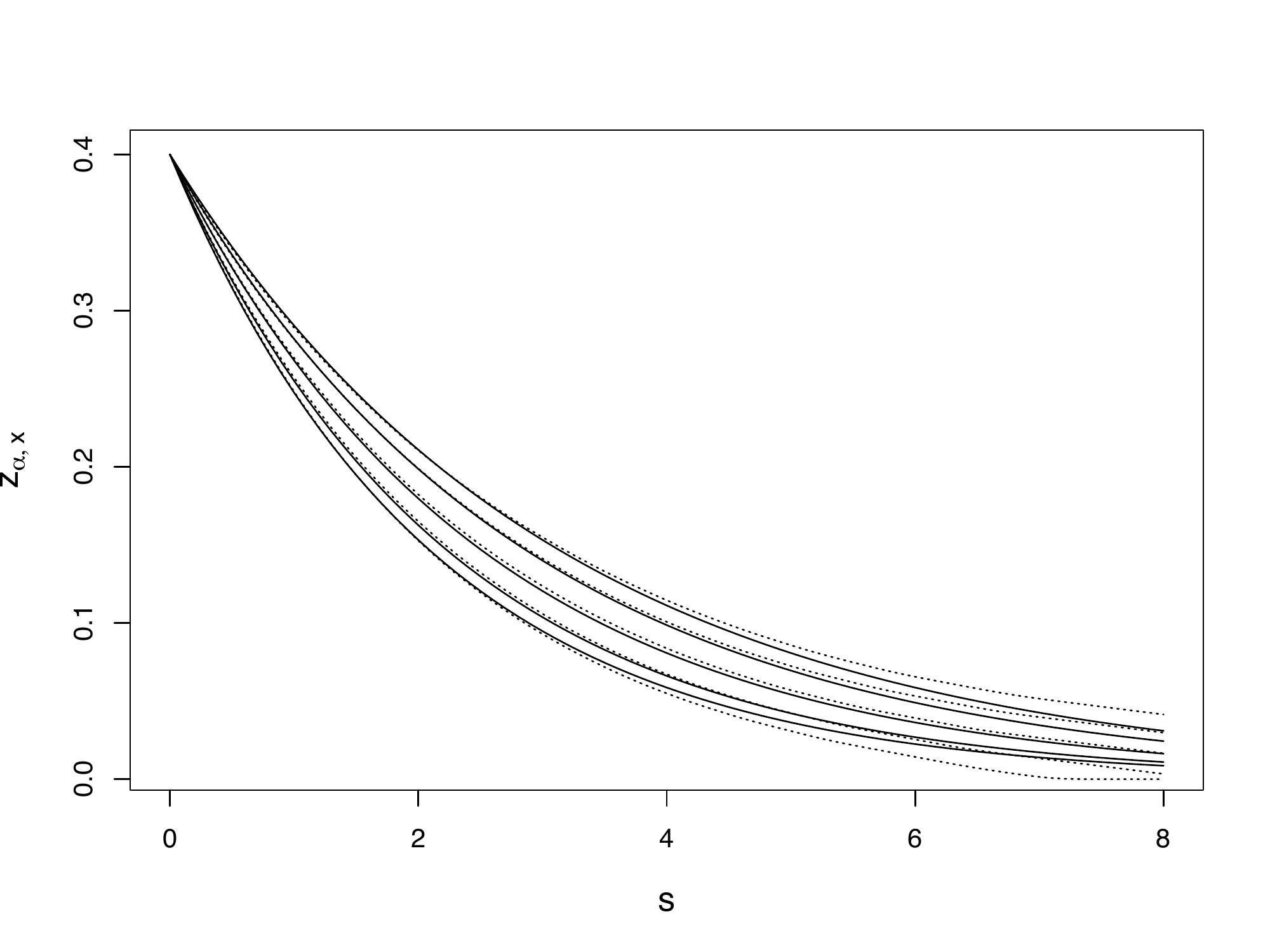}
  \end{subfigure}
  \begin{subfigure}[H]{.45\textwidth}
    \includegraphics[width = \textwidth]{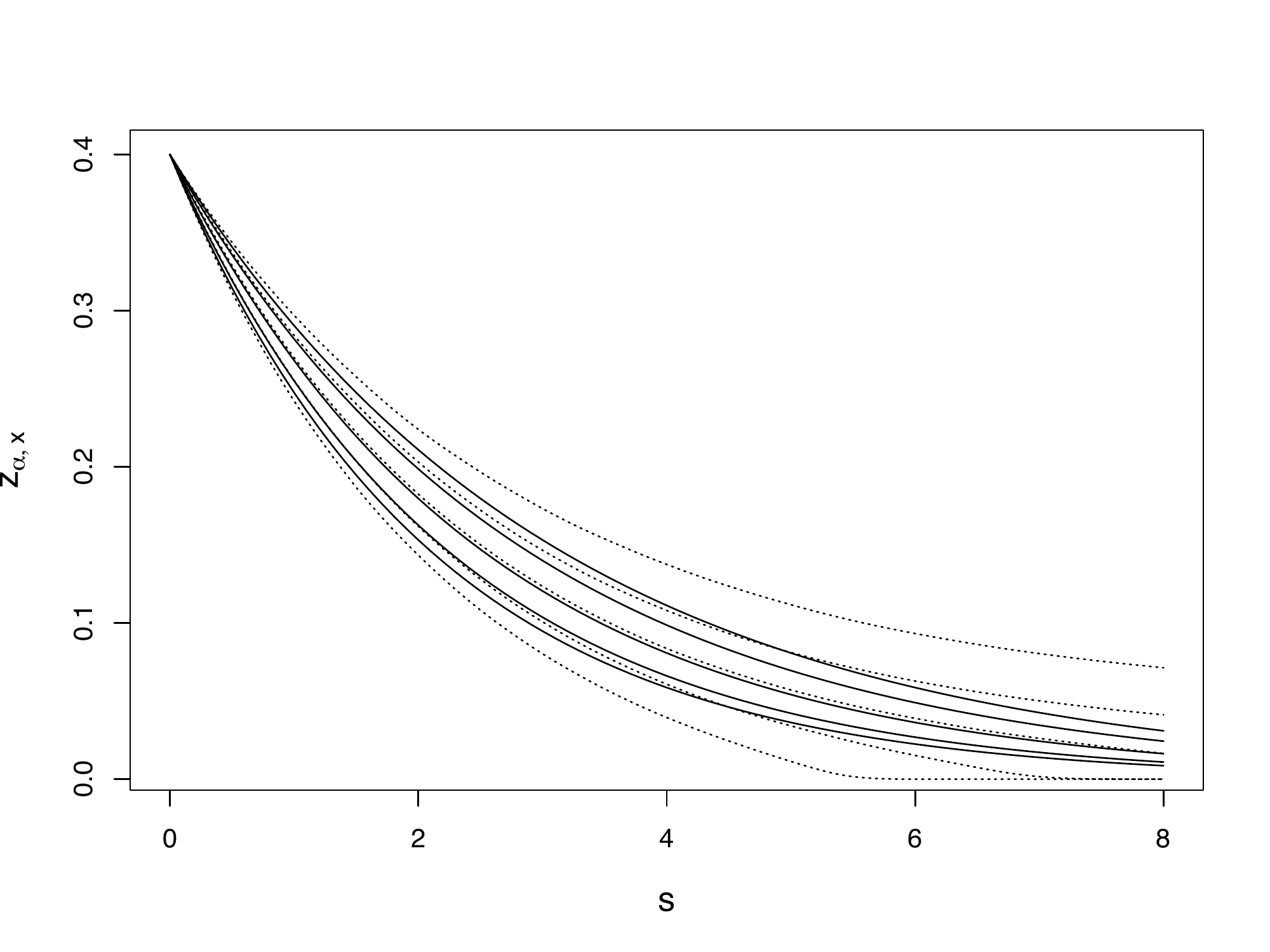}
  \end{subfigure}
  \begin{subfigure}[H]{.45\textwidth}
    \includegraphics[width = \textwidth]{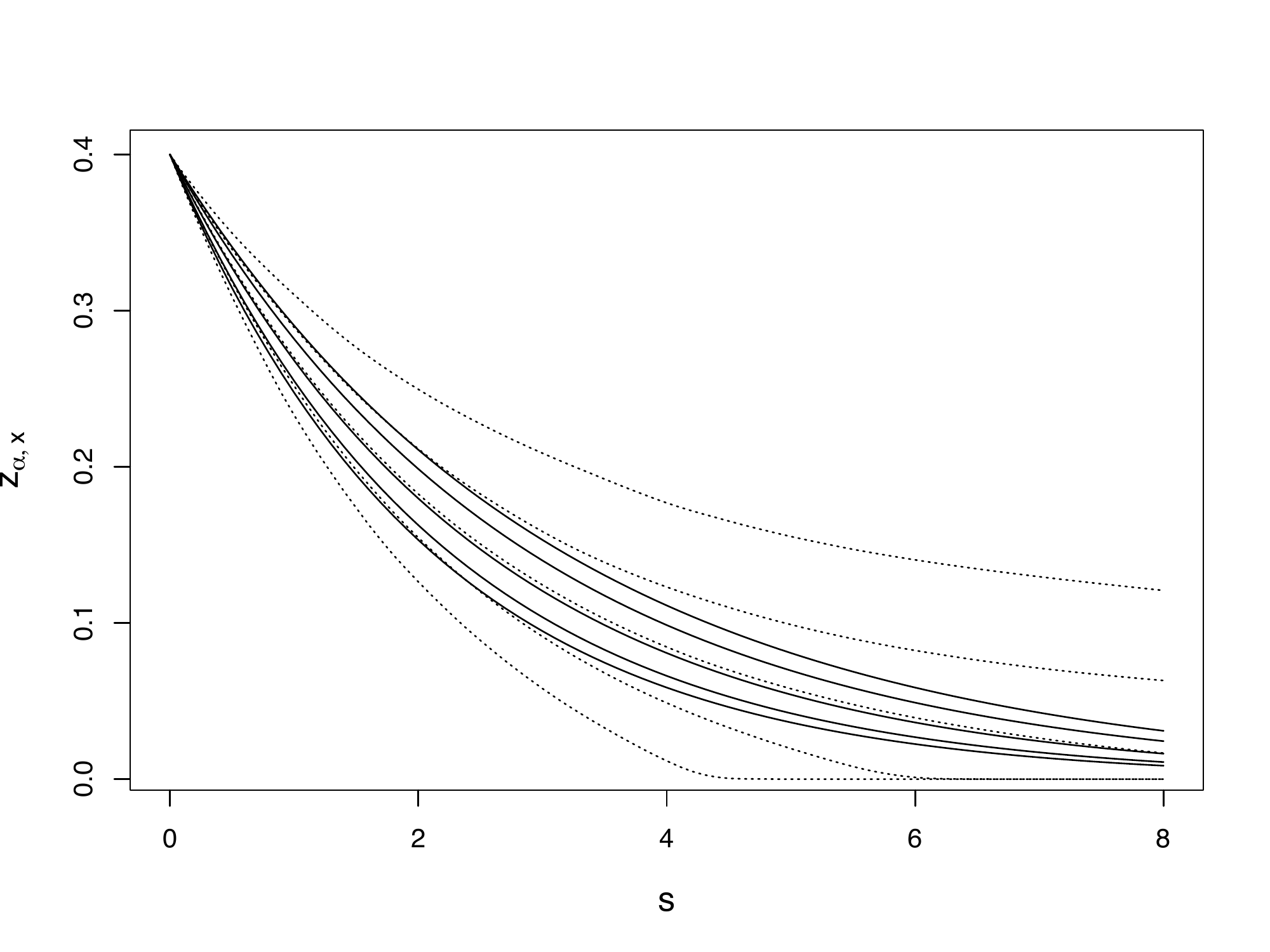}
  \end{subfigure}
    \caption{Average of 1000 simulation replications, each with a sample size of 300, where longitudinal measurements were taken with no noise (top left), noisy with $\sigma = .001$ (top right), noisy with $\sigma = .005$ (bottom left), and noisy with $\sigma = .01$ (bottom right). True quantiles are shown displayed with solid lines; estimates are shown with dotted lines.}
    \label{fig:expAvgd}
  \end{center}
\end{figure}

\begin{figure}[H]
  \begin{center}
    \includegraphics[height = 3.5in]{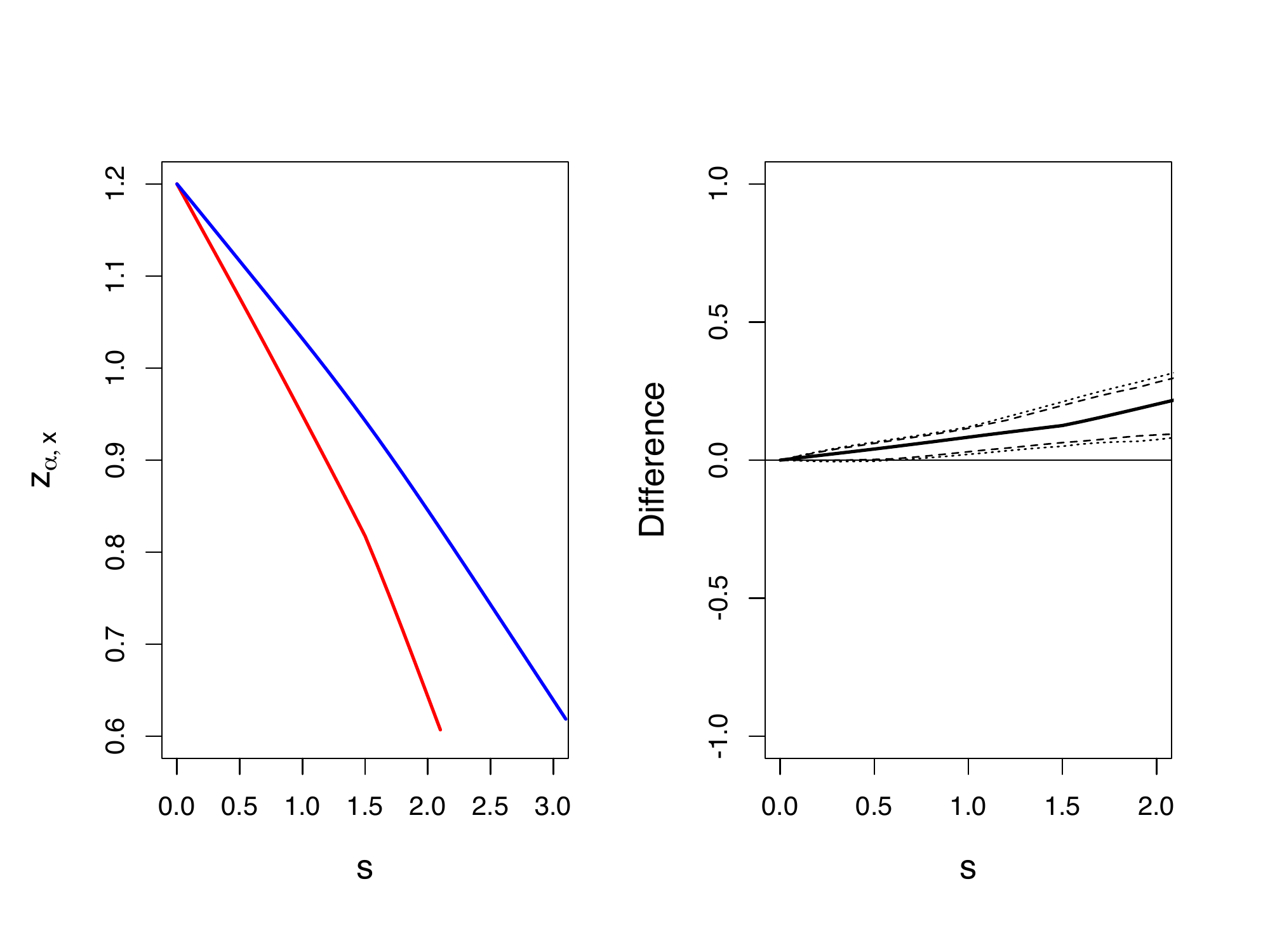}
    \caption{A low quantile comparison between normal and impaired subjects starting from $x = 1.2$. Here $\alpha = .05$ and 90\% and 95\% pointwise bootstrap confidence bands are included for the difference between normal and impaired subjects.}
    \label{fig:comparison05Plot}
  \end{center}
\end{figure}

\begin{figure}[H]
  \begin{center}
    \includegraphics[height = 2.2in]{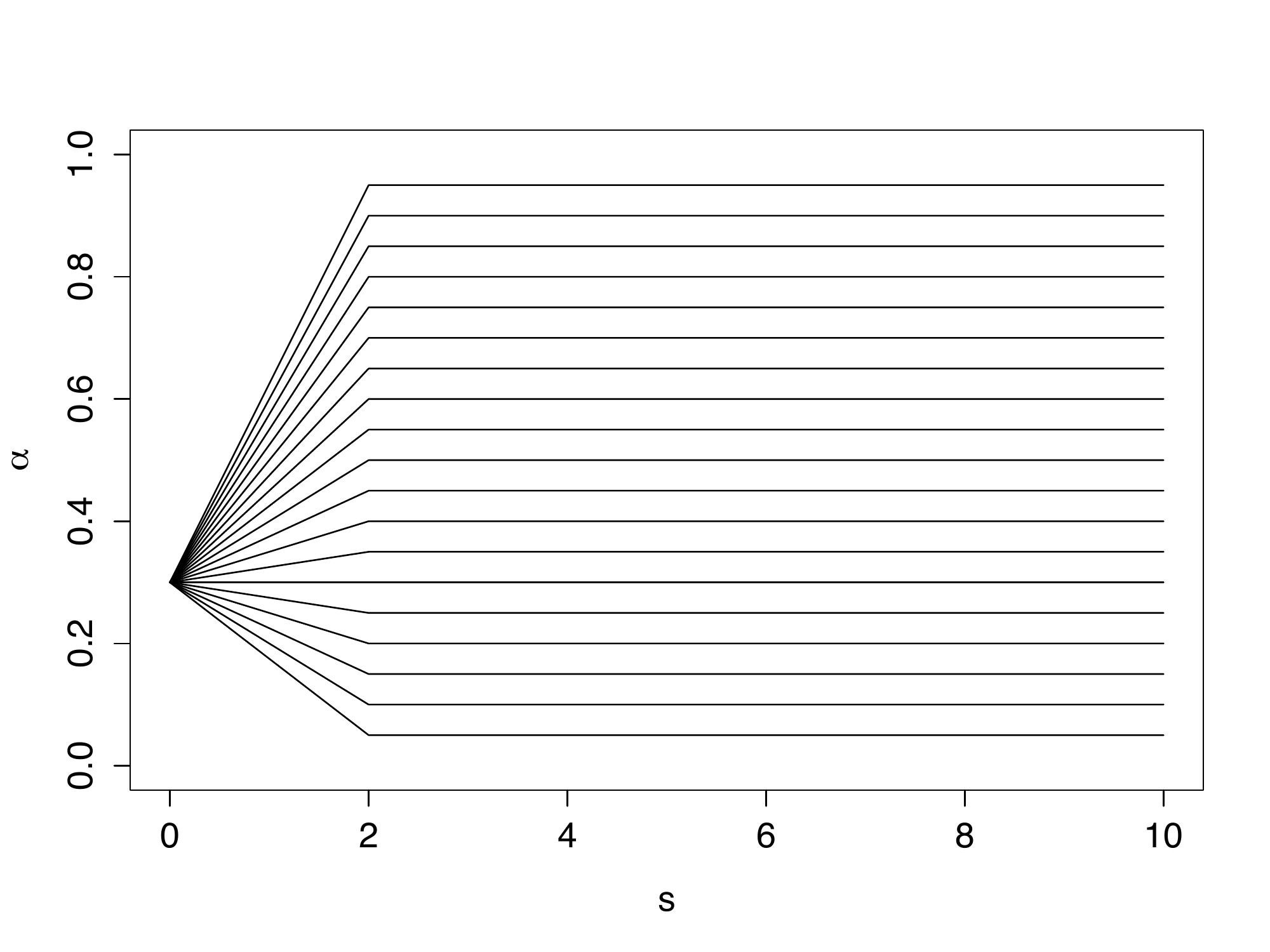}
    \caption{A demonstration of $\alpha(s)$ for $\alpha \in \{.05, \dots, .95\}$, starting from $\alpha^* = 0.3$ and with $S^* = 2$.}
    \label{fig:alphaPlot}
  \end{center}
\end{figure}
\ed
\begin{thebibliography}{26}
\newcommand{\enquote}[1]{``#1''}
\expandafter\ifx\csname natexlab\endcsname\relax\def\natexlab#1{#1}\fi

\bibitem[Abramson and M\"uller(1994)]{Abramson1994}
Abramson, I. and M\"uller, H.-G. (1994),
\newblock \enquote{Estimating direction fields in autonomous equation models, with an
  application to system identification from cross-sectional data,}
\newblock \textit{Biometrika}, 81, 663--672.

\bibitem[Albert et~al.(2011)Albert, DeKosky, Dickson, Dubois, Feldman, Fox,
  Gamst, Holtzman, Jagust, Petersen, Snyder, Carrillo, Thies, and
  Phelps]{Albert2011}
Albert, M.~S., DeKosky S.~T., Dickson D., Dubois, B., Feldman, H.~H. Feldman, Fox, N.~C., Gamst, A., Holtzman, D.~M., Jagust, W.~J., Petersen, R.~C., Snyder, P.~J., Carillo, M.~C., Thies, B., Phelps, C.~H (2011),
\newblock \enquote{The diagnosis of mild cognitive impairment due to Alzheimer's
  disease: recommendations from the National Institute on Aging-Alzheimer's Association
  workgroups on diagnostic guidelines for Alzheimer's disease,}
\newblock \textit{Alzheimer's \& Dementia : The Journal of the Alzheimer's
  Association}, 7, 270--279.

\bibitem[Baskin-Sommers et~al.(2015)Baskin-Sommers, Waller, Fish, and Hyde]{acc2}
Baskin-Sommers, A.R., Waller, R., Fish, A.M., Hyde, L.W. (2015),
\newblock \enquote{Callous-unemotional traits trajectories interact with earlier conduct problems and executive control to predict violence and substance use among high risk male adolescents,}
\newblock \emph{Journal of Abnormal Child Psychology}, 43,
  1529--1541.

\bibitem[Brault et~al.(2011)]{acc5}
Brault, M.C., Meuleman, B., Bracke, P. (2011),
\newblock \enquote{Depressive symptoms in the Belgian population: disentangling age and cohort effects,}
\newblock \emph{Social Psychiatry and Psychiatric Epidemiology}, 47,
  903--915.

\bibitem[Brumback and Bracke(1998)]{brum:98}
Brumback, B., Rice, J. (1998),
\newblock \enquote{Smoothing spline models for the analysis of nested and crossed samples of curves,}
\newblock \emph{Journal of the American Statistical Association}, 93,
  961--994.

\bibitem[Coffey et~al.(2014)]{coff:14}
Coffey, N., Hinde, J., Holian, E. (2014),
\newblock \enquote{Clustering longitudinal profiles using P-splines and mixed effects models applied to time-course gene expression data,}
\newblock \emph{Computational Statistics and Data Analysis}, 71,
  14--29.

\bibitem[Delaigle and Hall(2013)]{Delaigle2013}
Delaigle, A. and Hall P. (2013),
\newblock \enquote{Classification using censored functional data,}
\newblock \emph{Journal of the American Statistical Association}, 108,
  1269--1283.


\bibitem[Delaigle and Hall(2016)]{Delaigle2016}
Delaigle, A. and Hall P. (2016),
\newblock \enquote{Approximating fragmented functional data by segments of Markov chains,}
\newblock \emph{Biometrika}, 103, 779--799.


\bibitem[Dockery et~al.(1983)]{Dockery1983}
Dockery, D.~W., Berkey, C.~S., Ware, J.~H., Speizer, F.~E., Ferris, B.~G. (1983),
\newblock \enquote{Distribution of FVC and FEV1 in children 6 to 11 years old,}
\newblock \emph{American Review of Respiratory Disease}, 128,
405--412.

\bibitem[Fan and Gijbels(1996)]{Fan1996}
Fan, J. and Gijbels, I. (1996),
\newblock \emph{Local polynomial modelling and its applications,}
\newblock Chapman \& Hall/CRC.

\bibitem[Fan et~al.(1996)]{quant:fan}
Fan, J., Yao, Q. and Tong, H. (1996),
\newblock \enquote{Estimation of conditional densities and sensitivity measures in nonlinear dynamical systems,}
\newblock \emph{Biometrika}, 83, 189--206.

\bibitem[Ferraty et~al.(2006)Ferraty, Laksaci, and Vieu]{Ferraty2006}
Ferraty, F., Laksaci, A., Vieu, P. (2006),
\newblock \enquote{Estimating some characteristics of the conditional distribution in
  nonparametric functional models,}
\newblock \emph{Statistical Inference for Stochastic Processes}, 9,
  47--76.

\bibitem[Ford et~al.(2012)Ford, Hurd, Jagers, and Sellers]{acc1}
Ford, K., Hurd, N., Jagers, R., Sellers, R. (2012),
\newblock \enquote{Caregiver experiences of discrimination and African American adolescents' psychological health over time,}
\newblock \emph{Child Development}, 84,
  485--499.

\bibitem[Galbraith et~al.(2014)Galbraith, Bowden, and Mander]{Galbraith2014}
Galbraith, S., Bowden, J., Mander, A. (2014),
\newblock \enquote{Accelerated longitudinal designs: An overview of modelling, power, costs and handling missing data,}
\newblock \emph{Statistical Methods in Medical Research}, 2014, 0962280214547150.

\bibitem[Galla et~al.(2014)Galla, Wood, Tsukayama, Har, Chiu, Langer]{acc4}
Galla, B., Wood, J., Tsukayama, E., Har, K., Chiu, A., Langer, D. (2014),
\newblock \enquote{A longitudinal multilevel model analysis of the within-person and between-person effect of effortful engagement and academic self-efficacy on academic performance,}
\newblock \emph{Journal of School Psychology}, 52,
  295--308.

\bibitem[Gragg(1965)]{Gragg2006}
Gragg, W.B.(1965),
\newblock \enquote{On extrapolation algorithms for ordinary initial value problems,}
\newblock \emph{Journal of the Society for Industrial and Applied Mathematics: Series B, Numerical Analysis}, 2,
  384--403.

\bibitem[Guo(2004)]{Guo:04:1}
Guo, W. (2004),
\newblock \enquote{Functional data analysis in longitudinal settings using smoothing splines,}
\newblock \emph{Statistics in Medical Research}, 13,
  49--62.

\bibitem[Hall et~al.(1999)Hall, Wolff, and Yao]{Hall1999}
Hall, P., Wolff, R.~C., Yao, Q. (1999),
\newblock \enquote{Methods for estimating a conditional distribution function,}
\newblock \emph{Journal of the American Statistical Association}, 94,
  154--163.

\bibitem[Hall and M\"uller(2003)]{quant:hall}
Hall, P. and M\"uller, H.-G. (2003),
\newblock \enquote{Order-preserving nonparametric regression, with applications to conditional distribution and quantile function estimation,}
\newblock \emph{Journal of the American Statistical Association}, 98, 598--608.

\bibitem[Hansen(2008)]{Hansen2008}
Hansen, B.~E. (2008)
\newblock \enquote{Uniform convergence rates for kernel estimation with dependent data,}
\newblock \textit{Econometric Theory}, 24, 726--748.

\bibitem[Horrigue and Sa\"id(2011)]{Horrigue2011}
Horrigue, W. and Sa\"id, E.~O. (2011),
\newblock \enquote{Strong uniform consistency of a nonparametric estimator of a
  conditional quantile for censored dependent data and functional regressors,}
\newblock \emph{Random Operators and Stochastic Equations}, 19, 131--156.

\bibitem[Ioannides and Matzner-Lober(2009)]{Ioannides2009}
Ioannides, D.A. and Matzner-Lober, E. (2009),
\newblock \enquote{Regression quantiles with errors-in-variables,}
\newblock \emph{Journal of Nonparametric Statistics}, 21, 1003--1015.

\bibitem[Jiang and Wang(2011)]{jian:11}
Jiang, C.-R., Wang, J.-L. (2011),
\newblock \enquote{Functional single index model for longitudinal data,}
\newblock \emph{The Annals of Statistics}, 39,
  362--388.

\bibitem[Koenker(2005)]{quant:koenker1}
Koenker, R. (2005),
\newblock \emph{Quantile Regression,}
\newblock Cambridge University Press.

\bibitem[Koenker and Bassett(1978)]{quant:koenker2}
Koenker, R. and Bassett, G. (1978),
\newblock \enquote{Regression quantiles,}
\newblock \emph{Econometrika}, 46, 33--50.

\bibitem[Koenker et~al.(1994)]{quant:koenker3}
Koenker, R., Ng, P., and Portnoy, S. (1994),
\newblock \enquote{Quantile smoothing splines,}
\newblock \emph{Biometrika}, 81, 673--680.

\bibitem[{Kraus(2015)}]{Kraus2014}
Kraus, D. (2015), \enquote{Components and completion of partially observed functional data,} \textit{Journal of the Royal Statistical Society: Series B (Statistical Methodology),} 77, 777-801.

\bibitem[Liebl and Kneip(2016)]{liebl:16}
Liebl, D., Kneip, A. (2016), \enquote{On the optimal reconstruction of partially observed functional data,} \textit{Preprint,} \url{http://www.dliebl.com/files/Liebl_Kneip_FDA_Pred_2016.pdf}.

\bibitem[Li et~al.(2007)]{quant:li}
Li, Y., Liu, Y., and Zhu, J. (2007),
\newblock \enquote{Quantile regression in reproducing kernel hilbert spaces,}
\newblock \emph{Journal of the American Statistical Association}, 102, 255--268.

\bibitem[Li and Racine(2008)]{Li2008}
Li, Q. and Racine, J.~S. (2008),
\newblock \enquote{Nonparametric estimation of conditional CDF and quantile function
  with mixed categorical and continuous data,}
\newblock \textit{Journal of Business and Economic Statistics}, 26, 423--434.

\bibitem[McKhann et~al.(2011)McKhann, Knopman, Chertkow, Hyman, Clifford
  R.~Jack, Kawas, Klunk, Koroshetz, Manly, Mayeux, c.~Mohs, Morris, Rossor,
  Scheltens, Carrillo, Thies, Weintraub, and Phelps]{McKhann2011}
McKhann, G.~M., Knopman, D.~S., Chertkow, H., Hyman, B.~T., Clifford, J., Jack, C.R. Jr., Kawas, C.~H., Klunk, W.~E., Koroshetz, W.~J., Manly, J.~J., Mayeux, R.,
Mohs, R.~C., Morris, J.~C., Rossor, M.~N., Scheltens, P., Carrillo, M.~C.,
Thies, B., Weintraub, S., Phelps, C.~H. (2011),
\newblock \enquote{The diagnosis of dementia due to Alzheimer's disease: recommendations from the National Institute on Aging-Alzheimer's Association
  workgroups on diagnostic guidelines for Alzheimer's disease,}
\newblock \emph{Alzheimer's \& Dementia : the Journal of the Alzheimer's
  Association}, 7, 263--269.

\bibitem[Mu and Gage(2011)]{Mu2011}
Mu, Y. and Gage, F. (2011)
\newblock \enquote{Adult hippocampal neurogenesis and its role in Alzheimer's disease,}
\newblock \textit{Molecular Neurodegeneration}, 6:85.

\bibitem[Mungas et~al.(2004)Mungas, Reed, Crane, Haan, and
  Gonz\'alez]{DMungas2004}
Mungas, D., Reed, B., Crane, P., Haan, M., Gonz\'alez H. (2004),
\newblock \enquote{Spanish and English Neuropsychological Assessment Scales (SENAS):
  further development and psychometric characteristics,}
\newblock \textit{Psychological Assessment}, 16, 347--359.

\bibitem[{Raudenbush and Chan(1992)}]{Raudenbush1992}
Raudenbush, S. and Chan, W.-S. (1992),
\newblock \enquote{Growth Curve Analysis in Accelerated Longitudinal Designs,}
\newblock \textit{Journal of Research in Crime and Delinquency}, 29, 387--411.

\bibitem[Rice(2004)]{rice:04}
Rice, J. (2004),
\newblock \enquote{Functional and longitudinal data analysis: Perspectives on smoothing,}
\newblock \textit{Statistica Sinica}, 631--647.

\bibitem[Rice and Wu(2001)]{rice:01}
Rice, J., Wu, C. (2001),
\newblock \enquote{Nonparametric mixed effects models for unequally sampled noisy curves,}
\newblock \textit{Biometrics}, 57, 253--259.

\bibitem[{Roussas(1969)}]{roussas1969}
Roussas, G. (1969),
\newblock \enquote{Nonparametric estimation of the transition distribution function of a Markov process}
\newblock \textit{The Annals of Mathematical Statistics}, 40, 1386--1400.

\bibitem[{Sabuncu et~al.(2011)}]{Sabuncu2011}
Sabuncu, M.R., Desikan, R.S., Sepulcre, J., Yeo, B.T., Liu, H., Schmansky, N.J., Reuter, M., Weiner, M.W., Buckner, R.L., Sperling, R.A., Fischl, B. (2011),
\newblock \enquote{The dynamics of cortical and hippocampal atrophy in Alzheimer's disease,}
\newblock \textit{Archives of Neurology}, 68, 1040--1048.

\bibitem[Samanta(1989)]{Samant1989}
Samanta, M. (1989),
\newblock \enquote{Non-parametric estimation of conditional quantiles,}
\newblock \textit{Statistics and Probability Letters}, 7, 407--412.

\bibitem[Sperling et~al.(2011)Sperling, Aisen, Beckett, Bennett, Craft, Fagan,
  Iwatsubo, Jack, Montine, Park, Reiman, Rowe, Siemers, Stern, Yaffe, Carrillo,
  Thies, Morrison-Bogorad, Wagster, and Phelps]{Sperling2011}
Sperling, R.~A., Aisen, P.~S., Beckett, L.~A., Bennett, D.~A., Craft, S., Fagan, A.~M., Iwatsubo, T., Jack, C.~R. Jr., Montine, T.~J., Park, D.~C., Reiman, E.~M.,
  Rowe, C.~C., Siemers, E., Stern, Y., Yaffe, K., Carrillo, M.~C., Thies, B.,
  Morrison-Bogorad, M., Wagster, M.~V., Phelps, C.~H. Phelps (2011),
\newblock \enquote{Toward defining the preclinical stages of Alzheimer's disease:
  recommendations from the National Institute on Aging-Alzheimer's
  Association workgroups on diagnostic guidelines for Alzheimer's disease,}
\newblock \textit{Alzheimer's \& Dementia : The Journal of the Alzheimer's
  Association}, 7, 280--292, 2011.

\bibitem[Stanik et~al.(2013)Stanik, McHale and Crouter]{acc3}
Stanik, C.E., McHale, S.M., Crouter, A.C. (2013),
\newblock \enquote{Gender dynamics predict changes in marital love among African American couples,}
\newblock \emph{Journal of Marriage and Family}, 75,
  795--807.

\bibitem[Staniswalis et~al.(1998)]{stan:98}
Staniswalis, J., Lee, J. (1998),
\newblock \enquote{Nonparametric regression analysis of longitudinal data,}
\newblock \emph{Journal of the American Statistical Association}, 93,
  1403--1418.

\bibitem[{Vittinghoff et~al.(1994)}]{Vittinghoff1994}
Vittinghoff, E., Malani, H.M., Jewell, N.P (1994),
\newblock \enquote{Estimating patterns of CD4 lymphocyte decline using data from a prevalent cohort of HIV infected individuals,}
\newblock \textit{Statistics in Medicine}, 13, 1101--1118.

\bibitem[{Wang(2003)}]{Wang2003}
Wang, N. (2003),
\newblock \enquote{Marginal nonparametric kernel regression accounting for within-subject correlation,}
\newblock \textit{Biometrika}, 90, 43--52.

\bibitem[{Wang et~al.(2005)}]{Wang2005}
Wang, N., Carroll, R.J., and Lin, X. (2005),
\newblock \enquote{Efficient semiparametric marginal estimation for longitudinal/clustered data,}
\newblock \textit{Journal of the American Statistical Association}, 100, 147--157.

\bibitem[{Wei and Carroll(2009)}]{Wei2009}
Wei, Y., Carroll, R.J. (2009),
\newblock \enquote{Quantile regression with measurement error,}
\newblock \textit{Journal of the American Statistical Association}, 104, 1129--1143.

\bibitem[Wei and He(2006)]{quant:wei}
Wei, Y. and He, X. (2006),
\newblock \enquote{Conditional growth charts,}
\newblock \emph{The Annals of Statistics}, 34, 2069--2097.

\bibitem[{Yao et~al.(2005)Yao, M\"{u}ller, and Wang}]{Yao2005}
Yao, F., M\"{u}ller, H.-G., and Wang, J.-L. (2005), \enquote{Functional data
  analysis for sparse longitudinal data,} \textit{Journal of the American
  Statistical Association}, 100, 577--590.

\bibitem[Yu and Jones(1998)]{quant:yu}
Yu, K. and Jones, M.C. (1998),
\newblock \enquote{Local linear quantile regression,}
\newblock \emph{Journal of the American Statistical Association}, 93, 228--237.

\end{thebibliography}
